\documentclass[12pt]{article}
\usepackage{amsmath,color}
\input amssym

\begin{document}
\def\prg#1{\medskip\noindent{\bf #1}}     \def\ra{\rightarrow}
\def\lra{\leftrightarrow}        \def\Ra{\Rightarrow}
\def\nin{\noindent}              \def\pd{\partial}
\def\dis{\displaystyle}          \def\inn{\,\rfloor\,}
\def\Lra{{\Leftrightarrow}}
\def\Leff{\hbox{$\mit\L_{\hspace{.6pt}\rm eff}\,$}}
\def\bull{\raise.25ex\hbox{\vrule height.8ex width.8ex}}
\def\ads{(A)dS}

\def\G{\Gamma}        \def\S{\Sigma}        \def\L{{\mit\Lambda}}
\def\D{\Delta}        \def\Th{\Theta}
\def\a{\alpha}        \def\b{\beta}         \def\g{\gamma}
\def\d{\delta}        \def\m{\mu}           \def\n{\nu}
\def\th{\theta}       \def\k{\kappa}        \def\l{\lambda}
\def\vphi{\varphi}    \def\ve{\varepsilon}  \def\p{\pi}
\def\r{\rho}          \def\Om{\Omega}       \def\om{\omega}
\def\s{\sigma}        \def\t{\tau}          \def\eps{\epsilon}
\def\nab{\nabla}      \def\btz{{\rm BTZ}}   \def\heps{\hat\eps}

\def\tG{{\tilde G}}   \def\tR{{\tilde R}}
\def\bA{{\bar A}}     \def\bF{{\bar F}}     \def\bG{{\bar G}}
\def\bt{{\bar\tau}}   \def\bg{{\bar g}}     \def\cT{{\cal T}}
\def\cL{{\cal L}}     \def\cM{{\cal M }}    \def\cE{{\cal E}}
\def\cH{{\cal H}}     \def\bcH{{\bar\cH}}   \def\hcH{\hat{\cH}}
\def\cK{{\cal K}}     \def\hcK{\hat{\cK}}   \def\tcL{{\tilde\cL}}
\def\cO{{\cal O}}     \def\hcO{\hat{\cal O}}
\def\cV{{\cal V}}     \def\bV{{\bar V}}
\def\cB{{\cal B}}     \def\heps{\hat\epsilon}

\def\bb{{\bar b}}     \def\bt{{\bar\t}}     \def\br{{\bar\r}}
\def\bpi{{\bar\pi}}   \def\bom{{\bar\om}}   \def\bphi{{\bar\phi}}
\def\bJ{{\bar J}}     \def\bl{{\bar\l}}     \def\tom{{\tilde\omega}}

\def\hR{{\hat R}}     \def\hpi{{\hat\pi}}   \def\hPi{{\hat\Pi}}

\def\bi{{\bar\imath}} \def\bj{{\bar\jmath}} \def \bk{{\bar k}}
\def\bm{{\bar m}}     \def\bn{{\bar n}}     \def\bl{{\bar l}}
\def\bT{{\overline{T\mathstrut}}}
\def\bR{{\overline{R\mathstrut}}}
\def\cR{{\cal R}}       \def\cA{{\cal A}}
\def\hA {{\hat A}}      \def\hB{{\hat B}}   \def\hR{{\hat R}{}}
\def\hOm{{\hat\Omega}}  \def\hL{{\hat\L}}

\def\tb{{\tilde b}}    \def\tA{{\tilde A}}  \def\tV{{\tilde V}}
\def\tT{{\tilde T}}    \def\tR{{\tilde R}}  \def\tv{{\tilde V}}
\def\tcA{{\tilde\cA}}  \def\ts{\tilde\s}    \def\tcV{\tilde\cV}

\def\irr#1#2{\hspace{0.5pt}{}^{#1}\hspace{-1.6pt}#2}
\def\mb#1{\hbox{{\boldmath $#1$}}}
\def\fT{\mb{T}}       \def\fR{\mb{R}}       \def\fV{\mb{V}}
\def\fA{\mb{\cA}}     \def\tot{{\rm tot}}   \def\can{{\rm c}}

\def\nb{\marginpar{\bf\Huge ?}}
\def\nn{\nonumber}
\def\be{\begin{equation}}             \def\ee{\end{equation}}
\def\ba#1{\begin{array}{#1}}          \def\ea{\end{array}}
\def\bea{\begin{eqnarray} }           \def\eea{\end{eqnarray} }
\def\beann{\begin{eqnarray*} }        \def\eeann{\end{eqnarray*} }
\def\beal{\begin{eqalign}}            \def\eeal{\end{eqalign}}
\def\lab#1{\label{eq:#1}}             \def\eq#1{(\ref{eq:#1})}
\def\bsubeq{\begin{subequations}}     \def\esubeq{\end{subequations}}
\def\bitem{\begin{itemize}}           \def\eitem{\end{itemize}}
\renewcommand{\theequation}{\thesection.\arabic{equation}}

\title{3D gravity with propagating torsion:
Hamiltonian structure of the scalar sector}

\author{M. Blagojevi\'c and B. Cvetkovi\'c\footnote{
        Email addresses: {\tt mb@ipb.ac.rs,
                                cbranislav@ipb.ac.rs}} \\
University of Belgrade, Institute of Physics,\\ P. O. Box 57,
11001 Belgrade, Serbia}
\date{\today}
\maketitle

\begin{abstract}

We study the Hamiltonian structure of the general parity-invariant model of
three-dimensional gravity with propagating torsion, with eight parameters
in the Lagrangian. In the scalar sector, containing scalar or pseudoscalar
modes with respect to ma\-xi\-mally symmetric background, the phenomenon
of constraint bifurcation is observed and analyzed. The stability of the
Hamiltonian structure under linearization is used to identify dynamically
acceptable values of parameters.
\end{abstract}

\section{Introduction}
\setcounter{equation}{0}

Models of three-dimensional (3D) gravity were introduced to help us in
clarifying highly complex dynamical behavior of the realistic
four-dimensional general relativity (GR). In the last three decades, they
led to a number of outstanding results \cite{x1}. However, in the early
1990s, Mielke and Baekler \cite{x2} proposed a new, non-Riemannian
approach to 3D gravity, based on the Poincar\'e gauge theory (PGT)
\cite{x3,x4,x5,x6}. In contrast to the traditional GR with an underlying
Riemannian geometry of spacetime, the PGT approach is characterized by a
Riemann--Cartan geometry, with both the curvature and the torsion of
spacetime as carriers of the gravitational dynamics. Thus, PGT allows
exploring the interplay between gravity and geometry in a more general
setting.

Three-dimensional GR with or without a cosmological constant, as well as
the Mielke--Baekler (MB) model, are topological theories without
propagating modes. From the physical point of view, such a degenerate
situation is certainly not quite realistic. In the context of Riemannian
geometry, this limitation is surmounted by two well-known models:
topologically massive gravity \cite{x7} and the Bergshoeff--Hohm--Townsend
massive gravity \cite{x8}. On the other hand, including propagating modes
in PGT is much more natural: it is achieved simply by using Lagrangians
quadratic in the field strengths \cite{x9,x10,x11,x12}.

Since the general parity-invariant PGT Lagrangian in 3D is defined by
eight arbitrary para\-me\-ters \cite{x11}, it is a theoretical challenge
to find out which values of the parameters are allowed in a viable theory.
Following the approach of Sezgin and Niewenhuizen \cite{x13},
Helay\"el-Neto et al. \cite{x10} used the \emph{weak-field approximation}
around the Minkowski background to analyze this issue in a
parity-violating version of PGT, and found a number of interesting
restrictions on the parameters. However, one should be very careful with
the interpretation of these results, since (i) it is not clear how the
trensition from Minkowski to (anti-)de Sitter [(A)dS] background might
influence the perturbative analysis, and (ii) the weak-field approximation
does not always lead to a correct identification of the physical degrees
of freedom. Regarding (ii), we note that the constrained Hamiltonian
method \cite{x14,x4} is best suited for analyzing dynamical content of
gauge field theories, respecting fully their \emph{nonlinear structure}.
As noticed by Chen et al. \cite{x15} and Yo and Nester \cite{x16}, it may
happen, for some ranges of parameters, that the canonical structure of a
theory (the number and/or type of constraints) is changed after
linearization in a way that affects its physical content, such as the
number of physical degrees of freedom. Based on the \emph{canonical
stability under linearization} as a criterion for an acceptable choice of
parameters, Shie et al. \cite{x17} were able to define a PGT cosmological
model that offers a convincing explanation of dark energy as an effect
induced by torsion. Recently, the Bergshoeff--Hohm--Townsend massive
gravity is found to be canonically unstable under linearization
\cite{x18,x19}.

In this paper, we use the constrained Hamiltonian formalism to study (a)
the phenomenon of ``constraint bifurcation" and (b) the stability under
linearization of the general parity-invariant PGT in 3D \cite{x11}, in
order to find out the parameter values that define consistent models of 3D
gravity with propagating torsion. Because of the complexity of the
Hamiltonian structure, we restrict our attention to the scalar sector,
with $J^P=0^+$ or $0^-$ modes, defined with respect to the (A)dS
background. Investigation of higher spin modes is left for a future study.

The paper is organized as follows. In Section 2, we review basic
Lagrangian aspects of the parity-invariant PGT in 3D. In Section 3, we
give a brief account of the weak-field approximation around the (A)dS
background, restricting our attention to the scalar sector, with $J^P=0^+$
or $0^-$. In Section 4, we analyze general aspects of the canonical
dynamics of PGT; in particular, we examine how, depending on certain
critical values of parameters, some extra primary constraints may appear
(if-constraints), leading to a significant effect on the Hamiltonian
structure. In Section 5, we analyze the canonical structure of the
spin-$0^+$ sector, including the ``constraint bifurcation" effects. Then,
the test of canonical stability under linearization is used to reveal
dynamically acceptable values of parameters. In Section 6, the same type
of analysis is carried out for the spin-$0^-$ sector. Section 7 is devoted
to concluding remarks, and appendices contain technical details.

Our conventions are as follows: the Latin indices $(i,j,k, ...)$ refer
to the local Lorentz frame, the Greek indices $(\mu,\nu,\l, ...)$ refer
to the coordinate frame, and both run over 0,1,2; the metric components
in the local Lorentz frame are $\eta_{ij} = (+,-,-)$; totally
antisymmetric tensor $\ve^{ijk}$ is normalized to $\ve^{012}= 1$.

\section{Lagrangian formalism}
\setcounter{equation}{0}

We begin our considerations by a short account of the Lagrangian formalism
for PGT. Assuming parity invariance, the dynamics of 3D gravity with
propagating torsion is determined by the gravitational Lagrangian
(density) $\tcL_G=b\cL_G$,
\bsubeq\lab{2.1}
\be
\cL_G=-aR-2\L_0+\cL_{T^2}+\cL_{R^2}\, ,
\ee
where $\L_0$ is a bare cosmological constant, $a=1/16\pi G $, and the
pieces quadratic in the field strengths read:
\bea
\cL_{T^2}&:=&\frac{1}{2}T^{ijk}\left(a_1{}^{(1)}T_{ijk}
           +a_2{}^{(2)}T_{ijk}+a_3{}^{(3)}T_{ijk}\right)\,,\nn\\
\cL_{R^2}&:=&\frac{1}{4}R^{ijkl}\left( b_4{}^{(4)}R_{ijkl}
        +b_5{}^{(5)}R_{ijkl}+b_6{}^{(6)}R_{ijkl}\right)\,,
\eea
where $^{(n)}T_{ijk}$ and $^{(n)}R_{ijkl}$ are irreducible components of
the torsion and the Riemann--Cartan curvature \cite{x11}. Since the Weyl
curvature vanishes in 3D, one can rewrite these expressions in the form
that is more practical for the canonical analysis:
\bea
\cL_{T^2}&=&
  T^{ijk}\left(\a_1T_{ijk}+\a_2T_{kji}+\a_3\eta_{ij}V_k\right)\,,\nn\\
\cL_{R^2}&=&R^{ij}\left(\b_1R_{ij}+\b_2R_{ji}
            +\b_3\eta_{ij}R\right)=:R^{ij}\cH_{ij}\, ,
\eea
\esubeq
Here, $V_k:=T^m{}_{mk}$, $R_{ij}:=R^m{}_{imj}$ is the Ricci tensor, $R$ is
the scalar curvature, and
\bea
&&\a_1=\frac{1}{6}(2a_1+a_3)\, ,\quad\a_2=\frac{1}{3}(a_1-a_3)\, ,
  \quad \a_3=\frac{1}{2}(a_2-a_1)\, .                      \nn\\
&&\b_1=\frac{1}{2}(b_4+b_5)\, ,\quad \b_2=\frac{1}{2}(b_4-b_5)\, ,
     \quad \b_3=\frac{1}{12}(b_6-4b_4)\, .                 \nn
\eea
We also introduce the covariant momenta $\cH_{ijk}=\pd\cL_{G}/\pd
T^{ijk}$ and $\cH_{ijkl}=\pd\cL_{G}/\pd R^{ijkl}$:
\bea
\cH_{ijk}&=&2\left(a_1{}^{(1)}T_{ijk}
       +a_2{}^{(2)}T_{ijk}+a_3{}^{(3)}T_{ijk}\right)       \nn\\
  &=&4\left(\a_1 T_{ijk}+\a_2 T_{[kj]i}
                        +\a_3\eta_{i[j}v_{k]}\right)\, ,   \nn\\
\cH_{ijkl}&=&
  -2a(\eta_{ik}\eta_{jl}-\eta_{jk}\eta_{il})+\cH'_{ijkl}\,,\nn\\
\cH'_{ijkl}&=&2\left(b_4{}^{(4)}R_{ijkl}
     +b_5{}^{(5)}R_{ijkl}+b_6{}^{(6)}R_{ijkl}\right)       \nn\\
  &=&2(\eta_{ik}\cH_{jl}-\eta_{jk}\cH_{il})-(k\lra l)\, .  \nn
\eea

General field equations for the PGT theory \eq{2.1} are given in
\cite{x11}. Without matter contribution, these equations, transformed to
the local Lorentz basis, take the form:
\bsubeq\lab{2.2}
\bea
&&\nab^m\cH_{imj}
  +\frac{1}{2}\cH_i{}^{mn}(-T_{jmn}+2\eta_{jm}V_n)
  -t_{ij}=0\, ,                                             \\
&&2aT_{kij}+2T^m{}_{ij}(\cH_{mk}-\eta_{mk}\cH)
  +4\nab_{[i}(\cH_{j]k}-\eta_{j]k}\cH)
  +\ve_{ijn}\ve^{mr}{_k}\cH_{mr}{^n}=0\, ,
\eea
\esubeq
where $\cH=\cH^k{_k}$, and $t_{ij}$ is the energy-momentum tensor of
gravity:
$$
t_{ij}:=\eta_{ij}\cL_G-T^{mn}{_i}\cH_{mnj}+2a\hR_{ji}
  -2(\hR^n{_i}\cH_{nj}-\hR_j{}^{nm}{_i}\cH_{nm})\, .
$$

Relying again on the vanishing of the Weyl curvature, one can express
Bianchi identities in terms of the Ricci tensor. In the local Lorentz
basis, these identities take the form:
\bea
&&\ve^{mnr}\nab_m T^i{}_{nr}+\ve^{rsn}T^i{}_{mn}T^m{}_{rs}
  +2\ve^{imn}R_{mn}=0\, ,                                  \nn\\
&&\nab_k G^{ki}-V_kG^{ki}=0\, ,
\eea
where $G_{ki}:=R_{ki}-\frac{1}{2}\eta_{ik}R$.

\section{Scalar excitations around (A)dS background}
\setcounter{equation}{0}

Particle spectrum of 3D gravity with torsion \eq{2.1} around the Minkowski
background $M_3$ is already known \cite{x10,x11}. Here, we wish to examine
the modification of this spectrum induced by transition to the (A)dS
background. This will help us to clarify the relation between the
canonical stability of the theory under linearization and its $M_3$ or
(A)dS particle spectrum. Our attention is restricted to the scalar sector,
with $J^P=0^+,0^-$ modes.

Maximally symmetric configuration of 3D gravity with torsion is defined by
the set of fields $\bar\phi=(\bar b^i{_\m},\bar A^{ij}{_\m})$, such that
\be
\bT_{ijk}=p\ve_{ijk}\,,\qquad
\bR^{ij}{}_{mn}
  =-q\left(\d^i{_m}\d^j{_n}-\d^i{_n}\d^j{_m}\right)\, ,    \lab{3.1}
\ee
where the parameters $p$ and $q$ define an effective cosmological
constant,
$$
\Leff:=q-\frac{p^2}{4}\, .
$$
In order for this configuration to be a solution of the field equations in
vacuum, the para\-me\-ters $p$ and $q$ have to satisfy the following
conditions \cite{x11}:
\bsubeq\lab{3.2}
\bea
&&p(a+qb_6+2a_3)=0\, ,                                     \lab{3.2a}\\
&&aq-\L_0+\frac{1}{2}p^2a_3-\frac{1}{2}q^2b_6=0\, .        \lab{3.2b}
\eea
\esubeq
In the weak-field approximation around $\bar\phi$, the gravitational
variables $\phi=(b^i{_\m},A^{ij}{_\m})$ take the form
$\phi=\bar\phi+\tilde\phi$. We use the convention that indices of the
linear excitations $\tilde\phi$ are changed by the background triad and/or
metric.

The analysis of the particle spectrum is based on the linearized field
equations. In the same approximation, the Bianchi identities read:
\bsubeq\lab{3.3}
\bea
&&\ve^{kmn}\bar\nabla_k\tilde T^i{}_{mn}
  -2p\tilde V^i+2\ve^{imn}\tR_{mn}=0\, ,                   \lab{3.3a}\\
&&\bar\nabla_k\tilde G^{ki}-q\tilde V^i=0\, .              \lab{3.3b}
\eea
\esubeq

\subsection{Spin-\mb{0^+} mode} 

Looking at the particle spectrum of the theory \eq{2.1} on the $M_3$
background, see section 3 in \cite{x11}, one finds that the spin-$0^+$
mode has a finite mass (and propagates) if
\be
a_2(b_4+2b_6)\ne 0\, .                                     \nn
\ee
In order to study the spin-$0^+$ mode, we adopt the following, somewhat
simplified conditions:
\bsubeq\lab{3.4}
\be
a_2,b_6\ne 0\, ,\qquad a_1=a_3=b_4=b_5=0\, .               \lab{3.4a}
\ee
In fact, this choice is not unique since the existence of a spin-$0^+$
mode can be realized, for instance, without requiring $b_4=0$. However,
our ``minimal" choice \eq{3.4a} greatly simplifies the calculations, and
moreover, one does not expect that any essential dynamical feature of the
spin-$0^+$ mode will be thereby lost, see \cite{x15,x16}. The
corresponding Lagrangian reads:
\be
\cL^+_G=-aR-2\L_0
      +\frac{1}{2}a_2V^k V_k+\frac{1}{12}b_6 R^2\,,        \lab{3.4b}
\ee
and the conditions \eq{3.2} reduce to
\be
p(a+qb_6)=0\, ,\qquad aq-\L_0-\frac{1}{2}q^2b_6=0\, .
\ee
\esubeq

Now, we are going to show that the Minkowskian conditions \eq{3.4a}
equally well define the spin-$0^+$ mode with respect to the (A)dS
background \eq{3.1}. We start by noting that, under the conditions
\eq{3.4a}, the linearized field equations \eq{2.2} read:
\bsubeq\lab{3.5}
\bea
&&(a+qb_6)\tG_{ji}+a_2\eta_{i[j}\bar\nab^k\tV_{k]}
  +\frac{b_6q}{3}\eta_{ij}\tR=0\, ,                        \\
&&(a+qb_6)\tT_{ijk}-\frac{pb_6}{6}\ve_{ijk}\tR +a_2\eta_{i[j}\tV_{k]}
  +\frac{b_6}{3}\eta_{i[j}\bar\nab_{k]}\tR=0\, ,
\eea
\esubeq
and their traces are
\bsubeq\lab{3.6}
\bea
&&-2a_2\bar\nabla_i\tilde V^i+(a-qb_6)\tR=0\, ,            \lab{3.6a}\\
&&(a+qb_6+a_2)\tilde V_k+\frac{b_6}{3}\bar\nabla_k\tR=0\, .\lab{3.6b}
\eea
\esubeq
In the generic case, by combining $\bar\nabla_k\bar\nabla^k$ of \eq{3.6a}
with $\bar\nabla^k$ of \eq{3.6b}, one obtains
\be
\left(\bar\nabla_i\bar\nabla^i+m_{0^+}^2\right)\s=0\, ,\qquad
m_{0^+}^2=\frac{3(a-qb_6)(a+qb_6+a_2)}{2a_2b_6}\, ,        \lab{3.7}
\ee
where $\s:=\bar\nabla_i\tV^i$.  Thus, the field $\s$ can be identified as
the spin-$0^+$ excitation with respect to the (A)dS background, the mass of
which is finite. In the limit of vanishing $q$, $m^2_{0^+}$ reduces to the
corresponding Minkowskian expression.

\subsection{Spin-\mb{0^-} mode} 

Similar analysis can be applied to the spin-$0^-$ excitation. We start
from the Minkowski\-an condition that the spin-$0^-$ mode has a finite
mass (and propagates) \cite{x11},
$$
(a_1+2a_3)b_5\ne 0\, .
$$
We describe dynamics of the spin-$0^-$ sector by the simplified conditions:
\bsubeq
\be
a_3,b_5\ne 0\, ,\qquad a_1=a_2=b_4=b_6=0\, .               \lab{3.8a}
\ee
The related Lagrangian has the form
\be
\cL^-_G=-aR-2\L_0+3a_3\cA^2+b_5R_{[ij]}R^{[ij]}\, ,        \lab{3.8b}
\ee
with $\cA=\ve^{ijk}T_{ijk}/6$, and the conditions \eq{3.2} reduce to
\be
p(a+2a_3)=0\, ,\qquad aq-\L_0+\frac{1}{2}p^2a_3=0\, .      \lab{3.8c}
\ee
\esubeq

Starting from the linearized field equations,
\bsubeq
\bea
&&a_3\ve_{ijk}\bar\nab^k\tilde{\cA}+a_3p\eta_{ij}\tilde\cA
   +\frac{4a_3}{3}p\ve_{(imn}\tilde t_{j)}{}_{mn}
   -a_3p\ve_{ijk}\tV^k+a\tilde G_{ji}+b_5q\tilde R_{[ij]}=0\,,\\
&&a\tT_{ijk}+pb_5\ve^n{}_{jk}\tR_{[ni]}
   +b_5\bar\nab_{[j}(\tR_{k]i}
   -\tR_{ik]})+2a_3\ve_{ijk}\tilde{\cA}=0\, .
\eea
\esubeq
the axial irreducible components of these equations read:
\bsubeq
\bea
&&a_3\bar\nab^i\tilde{\cA}-a_3p\tV^i
   -\frac{1}{2}(a-qb_5)\ve^{ijk}\tR_{jk}=0\, ,             \nn\\
&&(a+2a_3)\tilde\cA
  +\frac{1}{3}b_5\ve^{ijk}\bar\nabla_i\tR_{jk}=0\, .       \nn
\eea
\esubeq
Then, the divergence of the first equation combined with the second one
yields
\be
a_3\bar\nabla^i\bar\nabla_i\tilde\cA-pa_3\bar\nabla_i\tilde V^i
+\frac{1}{2}(a-qb_5)\frac{3(a+2a_3)}{b_5}\tilde\cA=0\, .   \lab{3.11}
\ee
Now, using the divergence of the first Bianchi identity \eq{3.3a} and the
commutator identity $[\bar\nabla_m,\bar\nabla_n]\tilde X_i
  =-p\ve_{mnk}\bar\nabla^k \tilde X_i -2q\eta_{i[m}\tilde X_{n]}$,
we find
$$
\s\equiv\bar\nabla_k\tV^k=-\frac{3}{2}p(a+2a_3)\tilde\cA=0\, ,
$$
as a consequence of \eq{3.8c}. Hence, \eq{3.11} implies
\be
\left(\bar\nabla_k\bar\nabla^k+m_{0^-}^2\right)\tilde\cA=0\, ,\qquad
m_{0^-}^2=\frac{3(a-qb_5)(a+2a_3)}{2a_3b_5}\, .
\ee
Thus, generically, $\tilde\cA$ can be identified as the spin-$0^-$
excitation with respect to the (A)dS background.  For $q=0$, $m^2_{0^-}$
takes the Minkowskian form.

\section{Hamiltonian structure}
\setcounter{equation}{0}

In this section, we analyze general features of the Hamiltonian structure
of 3D gravity with propagating torsion, defined by the Lagrangian
\eq{2.1}; see \cite{x4,x20}.

\subsection{Primary constraints} 

We begin our study by analyzing the primary constraints. The canonical
momenta corresponding to basic dynamical variables
$(b^i{_\mu},A^{ij}{}_\m)$ are $(\pi_i{^\m},\Pi_{ij}{^\m})$; they are given
by
\be
\pi_i{^\m}:=\frac{\pd\tcL}{\pd(\pd_0 b^i{_\m})}
           =b\cH_i{}^{0\m}\, ,                             \qquad
\Pi_{ij}{^\m}:=\frac{\pd\tcL}{\pd(\pd_0 A^{ij}{}_\m)}
              =b\cH_{ij}{}^{0\m}\, .                       \nn
\ee
Since the torsion and the curvature do not involve the velocities
$\pd_0b^i{_0}$ and $\pd_0 A^{ij}{_0}$, one obtains the so-called ``sure"
primary constraints
\be
\pi_i{^0}\approx 0\, ,\qquad \Pi_{ij}{^0}\approx 0\, ,
\ee
which are always present, independently of the values of coupling
constants. If the Lagrangian \eq{2.1} is singular with respect to some of
the remaining velocities $\pd_0 b^i{_\a}$ and $\pd_0 A^{ij}{_\a}$, one
obtains further primary constraints. The existence of these primary
``if-constraints" (ICs) is determined by the critical values of the
coupling constants.

\prg{The torsion sector.} The gravitational Lagrangian \eq{2.1} depends on
the time derivative $\pd_0 b^i{_\a}$ only through the torsion tensor,
appearing in $\cL_{T^2}$. It is convenient to decompose $T_{ijk}$ into
the parallel and orthogonal components with respect to the spatial
hypersurface $\S$ (see Appendix A),
$$
T_{ijk}=T_{i\bj\bk}+2T_{i[\bj\perp}n_{k]}=\fT_{ijk}+\cT_{ijk}\,,
$$
where $\fT_{ijk}:=T_{i\bj\bk}$ does not depend on velocities and the
unphysical variables $(b^i{_0},A^{ij}{_0})$, and $n_k$ is the normal to
$\S$. Now, by introducing the parallel gravitational momentum
$\hpi_i{^\bk}=\pi_i{^\a}b^k{_\a}$ ($\hpi_i{^\bk}n_k=0$), one obtains
\bsubeq\lab{4.2}
\bea
\hpi_{i\bk}&=&J\cH_{i\perp\bk}(T)\, ,                      \lab{4.2a}
\eea
where $J:=\det(b^\bi{_\a})$, and
\be
\cH_{i\perp\bk}=2\left[2\a_1T_{i\perp\bk}+\a_2(T_{\bk\perp i}
  -T_{\perp\bk i})+\a_3(n_iV_\bk-\eta_{i\bk}V_\perp)\right]\, .\nn
\ee
The linearity of $\cH_{ijk}(T)$ in the torsion tensor allows us to rewrite
\eq{4.2a} in the form
\be
\phi_{i\bk}:=\frac{\hpi_{i\bk}}{J}-\cH_{i\perp\bk}(\fT)
            =\cH_{i\perp\bk}(\cT) \, ,
\ee
\esubeq
where the ``velocities"  $T_{i\bj\perp}$ appear only on the right-hand
side. This system of equations can be decomposed into irreducible parts
with respect to the group of two-dimensional rotations in $\S$. Going over
to the parameters $a_1,a_2,a_3$, one obtains:
\bsubeq\lab{4.3}
\bea
&&\phi_{\perp\bk}\equiv\frac{\hpi_{\perp\bk}}J-(a_2-a_1)T^\bm{}_{\bm\bk}
    =(a_1+a_2)T_{\perp\perp\bk} \,,                        \lab{4.3a}\\
&&\irr{S}{\phi}\equiv\frac{{}^S\hpi}J
  =-2a_2T^\bm{}_{\bm\perp}\,,                              \\
&&\irr{A}{\phi}_{\bi\bk}\equiv\frac{{}^A\hpi_{\bi\bk}}J
  -\frac{2}{3}(a_1-a_3)T_{\perp\bi\bk}=
  -\frac{2}{3}(a_1+2a_3)T_{[\bi\bk]\perp}\,,               \\
&&\irr{T}{\phi}_{\bi\bk}\equiv\frac{{}^T\hpi_{\bi\bk}}J
  = -2a_1\irr{T}{T}_{\bi\bk\perp}\, ,
\eea
\esubeq
where $\irr{S}{\phi}$, $\irr{A}{\phi}_{\bi\bk}$ and
$\irr{T}{\phi}_{\bi\bk}$ are the trace (scalar), antisymmetric and
traceless-symmetric parts of $\phi_{\bi\bk}$ (Appendix A).

If the critical parameter combinations appearing on the right-hand sides
of Eqs. \eq{4.3} vanish, the corresponding expressions $\phi_K$ become
additional primary constraints, the primary ICs. After a suitable
reordering, the result of the analysis is summarized as follows:
\bitem
\item[--] For $a_2=0$, $a_1+2a_3=0$, $a_1+a_2=0$ and/or $a_1=0$,
the expressions $\irr{S}{\phi}$, $\irr{A}{\phi}_{\bi\bk}$,
$\phi_{\perp\bk}$ and/or $\irr{T}{\phi}_{\bi\bk}$ become primary ICs
(see Table 1 below).
\eitem

\prg{The curvature sector.} In order to examine how the gravitational
Lagrangian depends on the velocities $\pd_0 A^{ij}{_\a}$, we start with
the following decomposition of the curvature tensor:
$$
R_{ijmn}=R_{ij\bm\bn}+2R_{ij[\bm\perp}n_{n]}=\fR_{ijmn}+\cR_{ijmn}\,,
$$
where $\fR_{ijmn}:=R_{ij\bm\bn}$ does not depend on the ``velocities"
$R_{ij\perp\bk}$ and the unphysical variables. The parallel
gravitational momentum $\hPi_{ij}{^\bk}=:\Pi_{ij}{^\a}b^k{_\a}$
($\hPi_{ij}{^\bk}n_k=0$) is given as
\bsubeq\lab{4.4}
\bea
\hPi_{ij\bk}&=&J\cH_{ij\perp\bk}(R)\, ,                    \lab{4.4a}
\eea
where
\bea
\cH_{ij\perp\bk}&=&-4an_{[i}\eta_{j]\bk}
    +4n_{[i}\cH_{j]\bk}-4\eta_{[i\bk}\cH_{j]\perp}         \nn\\
 &=&4n_{[i}\eta_{j]\bk}\left(-a+2\b_3R\right)              \nn\\
 &&+4\b_1\left(n_{[i}R_{j]\bk}-\eta_{[i\bk}R_{j]\perp}\right)
  +4\b_2\left(n_{[i}R_{\bk j]}-\eta_{[i\bk}R_{\perp j]}\right)\,.\nn
\eea
Since the ``velocities" $R_{ij\perp\bk}$ are contained only in $\cR$, we
rewrite this equation as
\be
\Phi_{ij\bk}:=\frac{\hPi_{ij\bk}}J+4an_{[i}\eta_{j]\bk}
     -\cH'_{ij\perp\bk}(\fR)=\cH'_{ij\perp\bk}(\cR)\, .    \lab{4.4b}
\ee
\esubeq
The components of a tensor $X_{\perp\bi\bj}$ can be decomposed into the
trace, antisymmetric and symmetric-traceless piece (Appendix A). Such a
decomposition of \eq{4.4b} yields:
\bsubeq\lab{4.5}
\bea
&&\irr{S}{\Phi}_\perp\equiv\frac{\irr{S}{\hPi}_\perp}{J}+4a
  -\frac{2}{3}(b_6-b_4)R^{\bk\bn}{}_{\bk\bn}
  =\frac{2}{3}(b_4+2b_6)R^\bk{}_{\perp\bk\perp}\,,         \lab{4.5a}\\
&&\irr{A}{\Phi}_{\perp\bi\bj}\equiv
   \frac{\irr{A}{\hPi}_{\perp\bi\bj}}J+2b_5R^\bk{}_{[\bi\bj]\bk}
  =2b_5R_{[\bi\perp\bj]\perp}\,,\\
&&\irr{T}{\Phi}_{\perp\bi\bj}\equiv\frac{\irr{T}{\hPi}_{\perp\bi\bj}}{J}
  -b_4\left(2R_{(\bi\bk\bj)}{}^\bk
            -\eta_{\bi\bj}R^{\bm\bn}{}_{\bm\bn}\right)
  =b_4\left(2R_{(\bi\perp\bj)\perp}
            -\eta_{\bi\bj}R^\bk{}_{\perp\bk\perp}\right)\, .
\eea
For a tensor $X_{\bi\bj\bk}=-X_{\bj\bi\bk}$, the pseudoscalar
$(\ve^{\bi\bj\bk}X_{\bi\bj\bk})$ and  the symmetric-traceless piece
$(X_{\bi(\bj\bk)}-\text{traces})$ identically vanish. Hence, Eq. \eq{4.4b}
implies one more relation:
\be
\irr{V}{\Phi}^\bi\equiv
       \frac{\irr{V}{\hPi}^\bi}J-(b_4-b_5)R_{\perp\bk}{}^{\bi\bk}
       =(b_4+b_5)R^{\bi\bk}{}_{\perp\bk}\, ,
\ee
\esubeq
where $\irr{V}{X}^\bi=X^{\bi\bj}{_\bj}$ (Appendix A).

Thus, when the parameters appearing on the right-hand sides of \eq{4.5}
vanish, we have the additional primary constraints $\Phi_K$. Combining
these relations with those obtained in the torsion sector, one find the
complete set of primary ICs, including their spin-parity characteristics
($J^P$), as shown in Table 1.
\begin{center}
\doublerulesep 1.6pt
\begin{tabular}{l l l}
\multicolumn{3}{c}{Table 1. Primary if-constraints}        \\
\hline\hline\rule[-5.5pt]{0pt}{20pt}
Critical conditions & Primary constraints &$J^P$           \\
\hline\rule[-7pt]{0pt}{21pt}
$a_2=0$        &$\irr{S}{\phi}\approx 0$ &                 \\[-1.2ex]
~$b_4+2b_6=0$  &$\irr{S}{\Phi}_{\perp}\approx 0$
               & \raisebox{1.6ex}{$0^+$}                   \\
\hline\rule[-7pt]{0pt}{21pt}
$a_1+2a_3=0$   &$\irr{A}{\phi}_{\bi\bk}\approx 0$ &        \\[-1.2ex]
~$b_5=0$       &$\irr{A}{\Phi}_{\perp\bi\bk}\approx 0$
               & \raisebox{1.6ex}{$0^-$}                   \\
\hline\rule[-7pt]{0pt}{21pt}
$a_1+a_2=0$  &${\phi}_{\perp\bk}\approx 0$ &               \\[-1.2ex]
~$b_4+b_5=0$ &$\irr{V}{\Phi}_\bk\approx 0$
             & \raisebox{1.6ex}{$1$}                       \\
\hline\rule[-7pt]{0pt}{21pt}
$a_1=0$  &$\irr{T}{\phi}_{\bi\bk}\approx 0$ &              \\[-1.2ex]
~$b_4=0$ &$\irr{T}{\Phi}_{\perp\bi\bk}\approx 0$
         & \raisebox{1.6ex}{$2$}                           \\
\hline\hline
\end{tabular}
\end{center}
This classification has a noteworthy interpretation: whenever a pair of
the ICs with specific $J^P$ is absent, the corresponding dynamical mode is
liberated and becomes a \emph{physical degree of freedom} (DoF). Thus, for
$a_2(b_4+2b_6)\ne 0$, the spin-$0^+$ ICs are absent, and the related DoF
becomes physical. Similarly, $(a_1+2a_3)b_5\ne 0$ implies that the
spin-$0^-$ DoF becomes physical. The results obtained here refer to the
full nonlinear theory; possible differences with respect to the
perturbative analysis (Section 3) will be discuss in Sections 5 and 6.

\subsection{General form of the Hamiltonian} 

Once we know the complete set of the primary ICs, we can construct first
the canonical and then the total Hamiltonian. Being interested only in the
gravitational degrees of freedom, we disregard the matter contribution.

\prg{Canonical Hamiltonian.} In the absence of matter, the canonical
Hamiltonian (density) is defined by
$$
\cH_\can=\pi_i{^\a}\dot b^i{_\a}
      +\frac{1}{2}\Pi_{ij}{^\a}\dot A^{ij}{_\a}-b\cL_G\, .
$$
Using the lapse and shift functions $N$ and $N^\a$, defined in Appendix A,
one can rewrite $\cH_\can$ in the Dirac--ADM form \cite{x4,x20}:
\bsubeq\lab{4.6}
\be
\cH_\can=N\cH_\perp+N^\a\cH_\a
         -\frac{1}{2}A^{ij}{_0}\cH_{ij}+\pd_\a D^\a\,,
\ee
where
\bea
&&\cH_\perp=\hpi_i{^\bj}T^i{}_{\perp\bj}
  +\frac{1}{2}\hPi_{ij}{}^\bk R^{ij}{}_{\perp\bk}-J\cL_G
  -n^i\nab_\a\pi_i{^\a}\,,                                 \nn\\
&&\cH_\a=\pi_i{^\b}T^i{}_{\a\b}+\frac 12\Pi_{ij}{}^\b R^{ij}{}_{\a\b}
  -b^i{_\a}\nab_\b\pi_i{^\b}\,,                            \nn\\
&&\cH_{ij}=2\pi_{[i}{^\a}b_{j]\a}+\nab_\a\Pi_{ij}{}^\a\,,\nn\\
&&D^\a=b^i{_\a}\pi_i{^\a}+\frac 12\Pi_{ij}{^\a} A^{ij}{_\a}\,.
\eea
\esubeq

The canonical Hamiltonian is linear in the unphysical variables
$(b^i{_0},A^{ij}{_0})$, and $\cH_\perp$ is the only dynamical part of
$\cH_\can$. The ``velocities" $T^i{}_{\perp\bk}, R^{ij}{}_{\perp\bk}$
appearing in $\cH_\perp$ can be expressed in terms of the phase-space
variables, using Eqs. \eq{4.3} and \eq{4.5}. Explicit calculation is
simplified by separating the torsion and the curvature contributions in
$\cH_\perp$:
\bea
&&\cH_\perp=2\L_0J+\cH_\perp^T+\cH_{\perp}^R\, ,           \nn\\
&&\cH^T_\perp:=\hpi^{i\bj}T_{i\perp\bj}-J\cL_{T^2}
  -n^i\nab_\a\pi_i{^\a}\,,                                 \nn\\
&&\cH^R_\perp:=\frac{1}{2}\hPi^{ij\bk}R_{ij\perp\bk}
                 -J\cL_{R^2}+aJ\fR\, .
\eea
The torsion piece of $\cH_\perp$ turns out to have the form (Appendix
A)
\bsubeq\lab{4.8}
\bea
\cH_{\perp}^T&=&\frac{1}{2}J\phi^2-J\cL_{T^2}(\fT)
                -n^i\nab_\a\pi_i{^\a}\, ,                  \\
\phi^2&:=&\frac{\l(a_1+a_2)}{a_1+a_2}(\phi_{\perp\bi})^2
  +\frac{\l(a_2)}{2a_2}(\irr{S}{\phi})^2                      \nn\\
&&+\frac{3}{2}\frac{\l(a_1+2a_3)}{(a_1+2a_3)}(\irr{A}{\phi}_{\bi\bj})^2
  +\frac{\l(a_1)}{2a_1}(\irr{T}{\phi}_{\bi\bj})^2\, .
\eea
\esubeq
where $\l(x)$ is the singular function
$$
\dis\frac{\l(x)}{x}=\left\{
\ba{cc}
\dis\frac{1}{x}\,, & x\neq 0  \\[9pt]
0\,,& x=0
\ea\right.
$$
which takes care of the conditions under which ICs become true
constraints. Similar calculations for the curvature part yields
\bsubeq\lab{4.9}
\bea
\cH_{\perp}^R&=&\frac{1}{4}J\Phi^2-J\cL_{R^2}(\fR)+aJ\fR\,, \\
\Phi^2&:=&\frac{\l(b_5)}{b_5}(\irr{A}{\Phi}_{\perp\bj\bk})^2
   +\frac{\l(b_4)}{b_4}(\irr{T}{\Phi}_{\perp\bj\bk})^2        \nn\\
  &&+\frac{3}{2}\frac{\l(b_4+2b_6)}{b_4+2b_6}(\irr{S}{\Phi}_\perp)^2
    +2\frac{\l(b_4+b_5)}{b_4+b_5}(\irr{V}{\Phi}_\bi)^2\, .
\eea
\esubeq

\prg{Total Hamiltonian.} The total Hamiltonian is defined by the
expression
\bsubeq
\be
\cH_{\tot}=\cH_c+u^k{_0}\phi_k{^0}+\frac{1}{2}u^{ij}\Phi_{ij}{^0}
          +(u\cdot\phi)+(v\cdot\Phi)\, ,
\ee
where $u$'s and $v$'s are arbitrary multipliers and
$(u\cdot\phi)+(v\cdot\Phi)$ denotes the contribution of all the primary
ICs. Formally, the existence of ICs is regulated by the form of the
related multipliers; for instance, $u_{\perp\bk}$ is given as
$u_{\perp\bk}:=[1-\l(a_1+a_2)]u'_{\perp\bk}$, and so on. Using the
irreducible decomposition technique, we find:
\bea
&&(u\cdot\phi):=u^{\perp\bk}\phi_{\perp\bk}
   +\irr{T}{u}^{\bi\bk}\,\irr{T}{\phi}_{i\bk}
   +\irr{A}{u}^{\bi\bk}\,\irr{A}{\phi}_{i\bk}
   +\frac{1}{2}\irr{S}{u}\,\irr{S}{\phi}\, , \nn\\
&&(v\cdot\Phi):=
    \irr{T}{v}^{\perp\bi\bk}\,\irr{T}{\Phi}_{\perp\bi\bk}
   +\irr{A}{v}^{\perp\bi\bk}\,\irr{A}{\Phi}_{\perp\bi\bk}
   +\frac{1}{2}\irr{S}{v}^\perp\,\irr{S}{\Phi}_\perp
   +\irr{V}{v}^\bk\,\irr{V}{\Phi}_\bk\, .
\eea
\esubeq

\prg{Consistency conditions.} Having found the form of the total
Hamiltonian, we can now apply Dirac's consistency algorithm to the primary
constraints, $\dot\phi_K=\{\phi_K,H_\tot\}\approx 0$, where $H_\tot=\int
d^3 x\cH_\tot$ and $\{X,Y\}$ is the Poisson bracket (PB) between $X$ and
$Y$; then, the procedure continues with the secondary constraints, and so
on \cite{x20}. In what follows, our attention will be focused on the
scalar sector, with $J^P=0^+$ or $0^-$ modes.

\section{Spin-\mb{0^+} sector}
\setcounter{equation}{0}

As one can see from Table 1, the absence of two spin-$0^+$ constraints,
$\irr{S}{\phi}$ and $\irr{S}{\Phi}_\perp$, is ensured by the condition
$a_2(b_4+2b_6)\ne 0$, whereby the spin-$0^+$ degree of freedom becomes
physical. To study the dynamical content of this sector, we adopt the
relaxed conditions
\be
a_2,b_6\ne 0\, ,\qquad a_1=a_3=b_4=b_5=0\, ,               \lab{5.1}
\ee
which define the Lagrangian $\cL^+_G$ as in \eq{3.4b}.

\subsection{Hamiltonian and constraints} 

\prg{Primary constraints.} In the spin-$0^+$ sector \eq{5.1}, general
considerations of the previous section lead to the following conclusions:
the set of primary constraints is given by
\be\lab{5.2}
\ba{ll}
\pi_i{^0}\approx 0\,,\quad & \Pi_{ij}{^0}\approx 0\,,      \\[4pt]
\irr{A}{\phi}_{\bi\bj}
    :=\dis\frac{\irr{A}{\hpi}_{\bi\bj}}{J}\approx 0\,, &
          \irr{T}{\phi}_{\bi\bj}
          :=\dis\frac{\irr{T}{\hpi}_{\bi\bj}}{J}\approx 0\,,\\[6pt]
\irr{A}{\Phi}_{\perp\bi\bj}
  :=\dis\frac{\irr{A}{\hPi}_{\perp\bi\bj}}{J}\approx 0\,,\qquad &
          \irr{T}{\Phi}_{\perp\bi\bj}
  :=\dis\frac{\irr{T}{\hPi}_{\perp\bi\bj}}{J}\approx 0\, ,  \\[6pt]
\irr{V}{\Phi}_\bi:=\dis\frac{\irr{V}{\hPi}_{\bi}}{J}\approx 0\,,&
\ea
\ee
the dynamical part of the canonical Hamiltonian has the form
\bea
\cH_\perp&=&J\left[\frac{1}{2a_2}(\phi_{\perp\bk})^2
  +\frac{1}{4a_2}(\irr{S}{\phi})^2
  +\frac{3}{16b_6}{(\irr{S}{\Phi}_{\perp}})^2\right]       \nn\\
  &&-J\cL_G^+(\fT,\fR)-n_i\nab_\a\pi^{i\a}\, ,             \lab{5.3}
\eea
where $\phi_{\perp\bk},\irr{S}{\phi}$, and $\irr{S}{\Phi}_\perp$ are the
``generalized" momentum variables defined in \eq{4.3} and \eq{4.5}, and
the total Hamiltonian reads
\be
\cH_\tot=\cH_\can+\irr{A}{u}^{\bi\bj}{}\irr{A}{\phi}_{\bi\bj}
  +\irr{T}{u}^{\bi\bj}\irr{T}{\phi}_{\bi\bj}
  +\irr{A}{v}^{\bi\bj}\irr{A}{\Phi}_{\bi\bj}
  +\irr{T}{v}^{\bi\bj}\irr{T}{\Phi}_{\bi\bj}
  +\irr{V}{v}^\bi\irr{V}{\Phi}_{\bi}\, .
\ee

\prg{Secondary constraints.} The consistency conditions of the sure
primary constraints $\pi_i{^0}$ and $\pi_{ij}{^0}$ produce the secondary
constraints
\bsubeq
\be
\cH_\perp\approx 0\, ,\qquad \cH_\a\approx 0\, ,
\qquad \cH_{ij}\approx 0\, ,                               \lab{5.5a}
\ee
where
\bea
\cH_\a&\approx& \hpi_\perp{^\bi} T_{\perp\a\bi}
  -\frac{1}{2}\irr{S}\hpi\fV_\a+\frac 12\irr{S}\hPi_\perp R_{\perp\a}
  -b^i{_\a}\nab_\b \pi_i{^\b}\, ,                          \nn\\
\cH_{\bi\bk}&\approx&\frac{\irr{A}{\hpi}_{\bi\bk}}{J}
  +\frac{\irr{S}{\hPi}_\perp}{2J}T_{\perp\bi\bk}\,,        \nn\\
\cH_{\perp\bk}&\approx&
   \frac{\hpi_{\perp\bk}}{J}-\frac{\irr{S}{\hPi}_\perp}{2J}\fV_\bk
  +\nab_\bk\frac{\irr{S}{\hPi}_\perp}{2J}\,.               \lab{5.5b}
\eea
\esubeq

Going over to the (eight) primary ICs, $X_M=(\irr{A}{\phi}, \irr{T}{\phi},
\irr{A}{\Phi},\irr{T}{\Phi},\irr{V}{\Phi})$, we note that the only
nonvanishing PBs among them are
\bea
&&\{\irr{A}{\Phi}_{\perp\bi\bj},\irr{A}{\phi}^{\bm\bn}\}
  \approx-\frac{\irr{S}{\hPi}_\perp}{2J^2}
         \d_i{}^{[\bm}\d_j{}^{\bn]}\d\,,                   \nn\\
&&\{\irr{T}{\Phi}_{\perp\bi\bj},\irr{T}{\phi}^{\bm\bn}\}
  \approx\frac{\irr{S}{\hPi}_\perp}{2J^2}
         \d_{(\bi}{}^{(\bn}\d_{\bj)}{}^{\bm)}\d\,.         \lab{5.6}
\eea
As long as $\irr{S}{\hPi}_\perp\ne 0$, the constraints
$(\irr{A}{\phi},\irr{T}{\phi},\irr{A}{\Phi},\irr{T}{\Phi})$ are second
class (SC) \cite{x4,x20}, and their consistency conditions fix the values
of the corresponding multipliers $(\irr{A}{u},\irr{T}{u}, \irr{A}{v},
\irr{T}{v})$ in $\cH_\tot$. On the other hand, $\irr{V}{\Phi}$ commutes
with all the other primary constraints, but not with its own secondary
pair $\chi_\bi=\{\irr{V}{\Phi}_\bi,H_\tot\}$, see \cite{x20}. Using
$\chi_\bi\approx J^{-1}\{\irr{V}{\hPi}_\bi,H_{\tot}\}$ and
\bea
&&\{\irr{V}{\hPi}_\bi,\cH_{mn}\}\approx 0\, ,\qquad
  \{\irr{V}{\hPi}_\bi,\cH_\a\}\approx 0\, ,                \nn\\
&&\{\irr{V}{\hPi}_\bi,\cH_\perp\}\approx
  J\left[\frac{\phi_{\perp\bi}}{a_2}\left(
  \frac{a_2}{2}-\frac{\irr{S}{\hPi}_\perp}{4J}\right)
  +\frac{a_2}{2}\fV_\bi
  +\nab_\bi\frac{\irr{S}{\hPi}_\perp}{4J}\right]\, ,       \nn
\eea
one ends up with
\be
\chi_\bi:=\frac{\phi_{\perp\bi}}{a_2}\left(
  \frac{\irr{S}{\hPi}_\perp}{4J}-\frac{a_2}{2}\right)
  -\frac{a_2}{2}\fV_\bi-\nab_\bi\frac{\irr{S}{\hPi}}{4J}\,.\lab{5.7}
\ee
The only nonvanishing PB involving $\chi_\bi$ is
\bea
\{\chi_\bi,\irr{V}{\Phi}_\bk\}=\frac{2}{a_2 J}\eta_{\bi\bk}
 \frac{\irr{S}{\hPi}_\perp}{4J}\left(
       \frac{\irr{S}{\hPi}_\perp}{4J}-a_2\right)\d\, .     \lab{5.8}
\eea
Thus, for $\irr{S}{\hPi}_\perp(\irr{S}{\hPi}_\perp-4Ja_2)\ne 0$, both
$\chi_\bi$ and $\irr{V}{\Phi}_\bk$ are SC. Consequently, the consistency
condition of $\chi_\bi$ determines the multiplier $\irr{V}{v}^\bi$, which
completes the consistency algorithm.

If the kinetic energy density in the Hamiltonian \eq{5.3} is to be
positive definite (``no ghosts"), the coefficients of $(\irr{S}{\phi})^2$
and $(\irr{S}{\Phi}_\perp)^2$ should be positive:
\be
a_2>0\, ,\qquad b_6>0\, .
\ee
On the other hand, $(\phi_{\perp\bk})^2$ gives a negative definite
contribution, but it is an interaction term, as can be seen from \eq{4.3a}
and \eq{5.5b}.

\subsection{Constraint bifurcation} 

In the previous discussion, we identified the conditions for which all the
ICs, $X'_M=\left(X_M,\chi\right)$, are SC. To calculate the determinant of
the $10\times 10$ matrix $\D^+_{MN}=\{X'_M,X'_N\}$,
$$
\D^+\approx\left|\begin{array}{cccccc}
   0 & 0 & \{\irr{A}{\phi},\irr{A}{\Phi}\} & 0 & 0 & 0  \\
   0 & 0 & 0 & \{\irr{T}{\phi},\irr{T}{\Phi}\} & 0 & 0  \\
   -\{\irr{A}{\phi},\irr{A}{\Phi}\} & 0 & 0 & 0 & 0 & 0 \\
   0 & -\{\irr{T}{\phi},\irr{T}{\Phi}\} & 0 & 0 & 0 & 0 \\
   0 & 0 & 0 & 0 & 0 & \{\irr{V}{\phi},\chi\}  \\
   0 & 0 & 0 & 0 & -\{\irr{V}{\phi},\chi\} & 0  \\
       \end{array} \right|
$$
we use \eq{5.6} and \eq{5.8}, which leads to
\be
\D^+\sim \left(\frac{\irr{S}{\Pi}_\perp}{4J}\right)^{10}
         \left(\frac{\irr{S}{\Pi}_\perp}{4J}-a_2\right)^4\, .
\ee
Introducing a convenient notation
\be
W:=\frac{\irr{S}{\Pi}_\perp}{4J}\, ,
\ee
we see that $\D^+$ can vanish only on a set (of spacetime points) of
measure zero, defined by $W=0$ or $W-a_2=0$. In other words, the condition
\be
W(W-a_2) \ne 0                                             \lab{5.12}
\ee
is fulfilled \emph{almost everywhere} (everywhere except on a set of
measure zero). Thus, our previous discussion can be summarized by saying
that all of the ICs are SC almost everywhere; the related (generic)
classification of constraints is shown in Table 2.
\begin{center}
\doublerulesep 1.6pt
\begin{tabular}{lll}
\multicolumn{3}{c}{Table 2. Generic constraints
                            in the $0^+$ sector} \\
                                                      \hline\hline
\rule{0pt}{12pt}
& First class \phantom{x} & ~~Second class \phantom{x}  \\
                                                      \hline
\rule[-1pt]{0pt}{16pt}
Primary
  ~& $\pi_i{^0}$, $\Pi_{ij}{^0}$   & ~~$X_M$            \\
                                                      \hline
\rule[-1pt]{0pt}{19pt}
Secondary\phantom{x}
  ~& $\cH'_{\perp}$, $\cH'_\a$, $\cH'_{ij}$ & ~~$\chi_\bi$\\
                                                      \hline\hline
\end{tabular}
\end{center}
The Hamiltonian constraints $\cH'_\perp,\cH'_\a$ and $\cH'_{ij}$ are first
class (FC) \cite{x4,x20}; they are obtained from \eq{5.3} and \eq{5.5b} by
adding the contributions containing the determined multipliers. With
$N=18$, $N_1=12$ and $N_2=10$, the dimension of the phase space is given
as $N^*=2N-2N_1-N_2=2$. Thus, the theory exhibits a single Lagrangian DoF
almost everywhere.

However, the determinant $\D^+$, being a field-dependent object, may
vanish in some regions of spacetime, changing thereby the number and/or
type of constraints and the number of physical degrees of freedom, as
compared to the generic situation described in Table 2. This effect, known
as the phenomenon of \emph{constraint bifurcation}, can be fully
understood by analyzing the dynamical behavior of the two factors in
\eq{5.12}. Although the complete analysis can be carried out in the
canonical formalism, we base our arguments on the Lagrangian formalism, in
order to simplify the exposition (see Appendix B).

\prg{1.} Starting with the second factor,
\be
\Om:=W-a_2\approx -\left(a-\frac{1}{6}b_6R+a_2\right)\,,   \lab{5.13}
\ee
where we used \eq{4.5a} to clarify the geometric interpretation, one can
prove the relation
\be
-\Om V_k +2\pd_k\Om\approx 0\, ,                           \lab{5.14}
\ee
which implies that the behavior of $\Om$ is limited to the following two
options (Appendix B):
\bitem
\item[(a)] either $\Om(x)$ vanishes globally, on the whole spacetime
           manifold,\vspace{-8pt}
\item[(b)] or it does not vanish anywhere.
\eitem
Which of these two options is realized depends upon the initial conditions
for $\Om$; choosing them in accordance with (b) extends the generic
behavior of $\Om$, $\Om\ne 0$ almost everywhere, to the whole spacetime.
This mechanism is the same as the one observed in the spin-$0^+$ sector of
the 4-dimensional PGT; compare \eq{5.14} with equation (4.20) in
\cite{x16}$_1$.

\prg{2.} We now focus our attention to the first factor in \eq{5.12},
\be
W\approx-\left(a-\frac{1}{6}b_6R\right)\, .
\ee
It is interesting that a solution for the $W$-bifurcation ($W=0$) can be
found by relying on the solution for the $\Om$-bifurcation, which is based
on choosing $\Om\ne 0$ on the initial spatial surface $\S$. Indeed, the
choice $\Om>0$ on $\S$ implies
\bsubeq
\be
\Om>0\quad \mbox{globally}.
\ee
Then, since $\Om=W-a_2$ ($a_2$ is positive), we find
\be
W>a_2\quad \mbox{globally}.
\ee
\esubeq
Thus, with $\Om>0$ and $W>a_2$, the problem of constraint bifurcation simply
disappears. Note that geometrically, the condition $W>a_2$ represents a
restriction on the Cartan scalar curvature, $b_6R>6(a+a_2)$. An equivalent
form of this relation is obtained by using the identity $R=\tR-2\s$, where
$\tR$ is Riemannian scalar curvature.

Thus, with a suitable choice of the initial conditions, one can ensure the
generic condition $\D^+\ne 0$ to hold \emph{globally}, so that the
constraint structure of the spin-$0^+$ sector is described exactly as in
Table 2. Any other situation, with $W=0$ or $W-a_2=0$, would not be
acceptable---it would have a variable constraint structure over the
spacetime, the property that could not survive the process of linearization.

\subsection{Stability under linearization}\label{sec53} 

Now, we are going to compare the canonical structure of the full nonlinear
theory with its linear approximation around maximally symmetric
background.

In the linear approximation, the condition of canonical stability
\eq{5.12} is to be taken in the lowest order (zeroth) approximation. Using
$\bR=-6q$, it reduces to
\be
(a+qb_6)(a+qb_6+a_2)\neq 0\, .                             \lab{5.17}
\ee
The three cases displayed in Table 3 define characteristic sectors of the
linear regime (see Appendix C).
\begin{center}
\doublerulesep 1.6pt
\begin{tabular}{ccccl}
\multicolumn{5}{c}{
               Table 3. Canonical stability in the $0^+$ sector}\\
                                                      \hline\hline
\rule{-1pt}{16pt}
    & $a+qb_6$ &~$a+qb_6+a_2$ & DoF & stability       \\[2pt]
                                                      \hline
\rule[-1pt]{0pt}{16pt}
(a) & $\ne 0$  & $\ne 0$      & 1 & stable            \\[2pt]
                                                      \hline
\rule[-1pt]{0pt}{16pt}
(b) & $=0$     & $\ne 0$      & 0 & unstable          \\[2pt]
                                                      \hline
\rule[-1pt]{0pt}{16pt}
(c) & $\ne 0$  & $=0$         & 1 & stable*           \\[2pt]
                                                      \hline\hline
\end{tabular}
\end{center}

(a) When the condition \eq{5.17} is satisfied, the nature of the
constraints remains the same as in Table 2, and we have a single
Lagrangian DoF, the massive spin-$0^+$ mode.

(b) Here, all ICs become FC, but only six of them are independent. Thus,
$N_1 =12 + 6 = 18$, and with $N_2 = 0$, the number of DoF's is zero:
$N^*=36 -2\times 18 = 0$.

(c) In this case, $\tilde\chi_\bk$ is not an independent constraint, and
$\irr{V}{\tilde\Phi}_\bk$  is FC. As compared to (a), the number and type
of constraints is changed according to $N_1\rightarrow N_1+2$,
$N_2\rightarrow N_2-4$, but the number of DoF's remains one ($N^*=2$),
corresponding to the massless spin-$0^+$ mode.

The case when both $a+qb_6$ and $a+qb_6+a_2$ vanish is not possible, since
$a_2\ne 0$.

To clarify the case (c), we need a more detailed analysis. Consider first
the case (a), in which the constraint $\tilde\chi_\bi$, defined in
\eq{C.3}, is replaced by an equivalent expression,
$\tilde\chi{}'_\bi=\tilde\hpi_{\perp\bi}/{\bar J}$. Then, the pair of SC
constraints $(\irr{V}{\tilde\Phi}_\bk,\tilde\chi{}'_\bk)$, with the
related Dirac brackets, defines the reduced phase space $\tR(a)$. Next,
consider the case (c), where $\tilde\chi_\bi$ does not exist and
$\irr{V}{\tilde\Phi}_\bk$ is FC. Here, we can introduce a suitable gauge
condition associated to $\irr{V}{\tilde\Phi}_\bk$, given by
$\tilde\chi''_\bk=\tilde\hpi_{\perp\bi}/{\bar J}$. The pair
$(\irr{V}{\tilde\Phi}_\bk,\tilde\chi{}''_\bk)$ defines the reduced phase
space $\tR(c)$, which coincides with the reduced phase space $\tR(a)$,
subject to the additional condition $a+qb_6+a_2=0$. Thus, the ``massless"
nonlinear theory, defined by $a+qb_6+a_2=0$, is essentially (up to a gauge
fixing) stable under the linearization. The star symbol in Table 3
(stable*) is used to remind us of this gauge fixing condition.

For the $M_3$ background ($p=q=0$ and $a\ne 0$), the case (b) is not
possible.

\section{Spin-$\mb{0^-}$ sector}
\setcounter{equation}{0}

For $(a_1+2a_3)b_5\ne 0$, the constraints
$\irr{A}{\phi}_{\bi\bk},\irr{A}{\Phi}_{\perp\bi\bk}$ in Table 1 are
absent, and the spin-$0^-$ mode becomes a physical degree of freedom.
Here, we study canonical features of the spin-$0^-$ sector by using the
specific conditions
\be
a_3,b_5\ne 0\, ,\qquad a_1=a_2=b_4=b_6=0\, ,               \lab{6.1}
\ee
which define the Lagrangian $\cL^-_G$ as in \eq{3.8b}.

\subsection{Hamiltonian and constraints} 

\prg{Primary constraints.} Applying the conditions \eq{6.1} to the general
considerations of Section 4, we find the following set of the primary
(sure and if-) constraints:
\be\lab{6.2}
\ba{ll}
\pi_i{^0}\approx 0\,,\quad & \Pi_{ij}{^0}\approx 0\,,      \\[4pt]
\irr{S}{\phi}:=\dis\frac{{}^S\hpi}J\approx 0\, , &
\irr{T}{\phi}_{\bi\bj}
     :=\dis\frac{\irr{T}{\hpi}_{\bi\bj}}J\approx 0\,,      \\[7pt]
\phi_{\perp\bi}:=\dis\frac{\hpi_{\perp\bi}}{J}\approx 0\,, & \\[7pt]
\irr{S}{\Phi}_\perp
     :=\dis\frac{\irr{S}{\hPi}_\perp}{J}+4a\approx 0\,,\qquad &
\irr{T}{\Phi}_{\perp\bi\bj}
     :=\dis\frac{\irr{T}{\hPi}_{\perp\bi\bj}}{J}\approx 0\,.
\ea
\ee
The dynamical part of the canonical Hamiltonian has the form
\bea
\cH_\perp&=&J\left[\frac{3}{8a_3}(\irr{A}{\phi}_{\bi\bj})^2
  +\frac{1}{4b_5}(\irr{A}{\Phi}_{\perp\bi\bj})^2
  +\frac{1}{2b_5}(\irr{V}{\Phi}_\bi)^2\right]                 \nn\\
  &&-J\cL_G^-(\fT,\fR)-n_i\nab_\a\pi^{i\a}\, ,
\eea
where $\irr{A}{\phi}_{\bi\bj},\irr{A}{\Phi}_{\perp\bi\bj}$ and
$\irr{V}{\Phi}_\bi$ are the ``generalized" momentum variables defined in
\eq{4.3} and \eq{4.5}, and the total Hamiltonian reads:
\be
\cH_T=\cH_c+\frac{1}{2}\irr{S}{u}\irr{S}{\phi}
 +\irr{T}{u}^{\bi\bj}\,\irr{T}{\phi}_{\bi\bj}
 +u^{\perp\bi}\phi_{\perp\bi}
 +\frac{1}{2}\irr{S}{v}^\perp\irr{S}{\Phi}_{\perp}
 +\irr{T}{v}^{\perp\bi\bj}\,\irr{T}{\Phi}_{\perp\bi\bj}\, .
\ee

\prg{Secondary constraints.} The consistency conditions of the primary
constraints $\pi_i{^0}$ and $\Pi_{ij}{^0}$ produce the usual secondary
constraints:
\bsubeq
\be
\cH_\perp\approx 0\,,\quad \cH_\a\approx 0\, ,
\qquad \cH_{ij}\approx 0\, ,                               \lab{6.5a}
\ee
where
\bea
&&\cH_\a\approx\irr{A}\hpi^{\bi\bj}T_{\bi\a\bj}
  +\irr{A}\hPi_{\perp\bi\bj}R_{\perp}{^\bi}{_\a}{^\bj}
  +R^{\bi\bj}{}_{\a\bj}\irr{V}\hPi_{\bi}
  -2aJR_{\perp\a}-b^i{_\a}\nab_\b \pi_i{^\b}\, ,           \nn\\
&&\cH_{\bi\bj}\approx aT_{\perp\bi\bj}
  +\frac{\irr{A}{\hpi}_{\bi\bj}}{2J}
  +\frac{\irr{V}{\Pi}_\bk}{2J}T^\bk{}_{\bi\bj}
  +\nab_{[\bi}\frac{\irr{V}{\hPi}_{\bj]}}{J}\, ,           \nn\\
&&\cH_{\perp\bi}\approx a\fV_\bi
 +\frac{\irr{A}{\hPi}_{\perp\bm\bn}}{2J}T^{\bm\bn}{}_\bi
 +\frac{\irr{V}{\hPi}^\bm}{2J}T_{\perp\bi\bm}
 +\frac{1}{2}\nab_\bm\frac{\irr{A}\hPi_{\perp\bi}{}^\bm}J\, .
\eea
\esubeq

Using the PB algebra between the primary ICs
$Y_M=(\irr{S}{\phi},\irr{T}{\phi},\phi_{\perp,\bk}, \irr{S}{\Phi},
\irr{T}{\Phi})$ (Appendix D), one finds that generically, for
$\irr{A}{\hpi}_{\bi\bk}\ne 0$, they are SC; their consistency conditions
result in the determination of the corresponding multipliers
$(\irr{S}{u},\irr{T}{u},u_{\perp,\bk}, \irr{S}{v},\irr{T}{v})$. Moreover,
the secondary constraints \eq{6.5a}, corrected by the contributions of the
determined multipliers, are FC, so that their consistency conditions are
trivially satisfied. Thus, in the generic case, the consistency algorithm
is completed at the level of secondary constraints.

The first two terms in $\cH_\perp$, proportional to the squares of
$\irr{A}{\phi}_{\bi\bk}$ and $\irr{A}{\Phi}_{\perp\bi\bk}$, describe the
contribution of the spin-$0^-$ mode to the kinetic energy density, see
Table 1. This contribution is positive definite for
\be
a_3>0\, ,\qquad b_5>0\, .
\ee
At the same time, the contribution of the third term, the square of
$\irr{V}{\Phi}_\bk$, becomes negative definite (``ghost"), which is a
\emph{serious problem} for the physical interpretation. As we shall see,
this is not the only problem.

\subsection{Constraint bifurcation} 

Based on the PB algebra of the (eight) primary ICs $Y_M$, we can now
calculate the determinant of the $8\times 8$ matrix
$\D^-_{MN}=\{Y_M,Y_N\}$ (Appendix D); the result takes the form
\be
\D^-\sim \irr{A}\hpi_{\bi\bj}\irr{A}\hpi^{\bi\bj}
  \left(\frac{4a^2}{J^2}+\frac{1}{8J^4}\irr{A}\hPi_{\perp\bm\bn}
        \irr{A}\hPi^{\perp\bm\bn}\right)^2\, .             \lab{6.7}
\ee
Since the second factor is always positive definite, $\D^-$ remains
different from zero only if
\be
\irr{A}{\hpi}_{\bm\bn}\ne 0\, .
\ee
This condition holds everywhere except on a set of measure zero, so that
$\D^-\ne 0$ almost everywhere. Thus, generically, the eight primary ICs
are SC, as shown in Table 4; the primes in $\cH'_\perp,\cH'_\a$ and
$\cH'_{ij}$ denote the presence of corrections induced by the determined
multipliers.
\begin{center}
\doublerulesep 1.8pt
\begin{tabular}{lll}
\multicolumn{3}{l}{\hspace{0pt}Table 4. Generic constraints in
                                         the $0^-$ sector} \\
                                                      \hline\hline
\rule{0pt}{12pt}
&~First class \phantom{x}&~ Second class \phantom{x}       \\
                                                      \hline
\rule[-1pt]{0pt}{16pt}
\phantom{x}Primary &~$\pi_i{^0}$, $\Pi_{ij}{^0}$ &~ $Y_M$ \\
                                                      \hline
\rule[-1pt]{0pt}{19pt}
\phantom{x}Secondary &~$\cH'_{\perp}$, $\cH'_\a$, $\cH'_{ij}$
           &~                                         \\
                                                      \hline\hline
\end{tabular}
\end{center}
Using $N=18$, $N_1=12$ and $N_2=8$, we find $N^*=2N-2N_1-N_2=4$.
Surprisingly, the theory exhibits \emph{two} Lagrangian DoF: one is the
expected spin-$0^-$ mode, and the other is the spin-1 ``ghost" mode,
represented canonically by $\irr{V}{\Phi}_\bk$.

In Appendix E, we analyze the nature of the critical condition
$\irr{A}{\hpi}_{\bm\bn}=0$. In the region of spacetime where it holds, we
find the phenomenom of constraint bifurcation: the number of DoF is
changed to zero. Although such a situation is \emph{canonically unstable
under linearization}, it is interesting to examine basic aspects of the
linearizad theory.

\subsection{Linearization} 

In the linearized theory, the term $\irr{A}{\bar\hpi}_{\bj\bk}$ in the
determinant $\D^-$ takes the form
\be
\irr{A}{\bar\hpi}_{\bj\bk}=-2a_3\ve_{\perp\bj\bk}p\, .
\ee
Hence, the canonical structure of the linearized theory crucially depends
on the value of the background parameter $p$, as shown in Table 5.

\begin{center}
\doublerulesep 1.6pt
\begin{tabular}{cccl}
\multicolumn{4}{c}{
               Table 5. Canonical instability in the $0^-$ sector}\\
                                                      \hline\hline
\rule{-1pt}{16pt}
       &           & DoF     & stability              \\[2pt]
                                                      \hline
\rule[-1pt]{0pt}{16pt}
$(\a)$ & $p\ne 0$  & 2       & stable almost everywhere \\[2pt]
                                                      \hline
\rule[-1pt]{0pt}{16pt}
$(\b)$ & $p=0$     & 1       & unstable               \\[2pt]
                                                      \hline\hline
\end{tabular}
\end{center}

$(\a)$ For $p\ne 0$ (Riemann--Cartan background, massless spin-$0^-$
mode), the determinant $\bar\D^-$ is positive definite, all the primary
ICs are SC, as in the generic sector of the full nonlinear theory, and
consequently, $N^*=4$. However, this is not true in the critical region
$\irr{A}{\hpi}_{\bi\bj}=0$, where $N^*=0$ and the theory is canonically
unstable.

$(\b)$ For $p=0$ (Riemannian background, massive or massless spin-$0^-$
mode), the situation is changed (Appendix F). First, the determinant
$\bar\D^-$ vanishes, since the primary IC $\tilde\phi_{\perp\bi}$
\emph{commutes} with itself, see \eq{D.1}. By calculating its consistency
condition (which was not needed for $p\ne 0$), one finds its secondary
pair $\tilde\chi_\bi$. Now, the PB of $\tilde\phi_{\perp\bi}$ with the
modified secondary pair $\tilde\chi'_\bi=\tilde\chi_\bi-\tilde\cH_\bi$
does not vanish. Thus, there are two SC constraints more than in the case
$(\a)$ so that $N^*=2$, and we have the canonical instability under
linearization.

Thus, in both cases $(\a)$ and $(\b)$, the theory canonically unstable.

\section{Concluding remarks}
\setcounter{equation}{0}

In this paper, we studied the Hamiltonian structure of the general
parity-invariant model of 3D gravity with propagating torsion, described
by the eight-parameter PGT Lagrangian \eq{2.1}. Because of the complexity
of the problem, we focused our attention on the scalar sector, containing
$J^P=0^+$ or $0^-$ modes with respect to maximally symmetric background.
By investigating fully nonlinear ``constraint bifurcation" effects as well
as the canonical stability under linearization, we were able to identify
the set of dynamically acceptable values of parameters for the spin-$0^+$
sector, as shown in Table 3. On the other hand, the spin-$0^-$ sector is
found to be canonically unstable for any choice of parameters, see Table
5. Transition from an (A)dS to Minkowski background simplifies the
results.

Further analysis involving higher spin sectors is left for future studies.

\section*{Acknowledgements}

It is a pleasure to thank Vladimir Dragovi\'c for a helpful discussion.
This work was supported by the Serbian Science Foundation under Grant No.
171031.

\appendix
\section{The 1+2 decomposition of spacetime}
\setcounter{equation}{0}

To derive the Dirac-ADM form of the Hamiltonian, it is convenient to pass
from the tetrad basis $\mb{h}_i=h_i{^\m}\pd_\m$ to the ADM basis
$(\mb{n},{\mb{h}}_\a)$, where $\mb{n}$ is the unit vector with
$n_k={h_k{^0}}/{\sqrt{g^{00}}}$, orthogonal to the vectors
$\mb{h}_\a=\pd_\a$ lying in the $x^0=$ const. hypersurface $\S$, see
\cite{x4,x20}.

\prg{(1)} Introducing the projectors on $\mb{n}$ and $\S$,
$(P_\perp)^i_k=n^in_k$ and $(P_{||})^i_k=\d^i_k-n^in_k$, any vector $V_k$
can be decomposed in terms of its normal and parallel projections:
\bea
&&V_k=n_kV_\perp+V_{\bk}\, ,                               \lab{A.1}\\
&&V_\perp:=n^k V_k\, ,\qquad V_\bk:=(P_{||})^i_kV_i
                                   =h_\bk{^\a}V_\a\, .     \nn
\eea
The decomposition of $V_0=b^k{_0}V_k$ in the ADM basis yields
$V_0=NV_\perp+N^\a V_\a$, where the lapse and shift functions $N$ and
$N^\a$, respectively, are linear in $b^k{_0}$:
\be
N:=n_k b^k{_0}\, ,\qquad N^\a:=h_\bk{^\a}b^k{_0}\, .       \lab{A.2}
\ee
The decomposition \eq{A.1} can be extended to any tensor field. Thus, a
second rank antisymmetric tensor $X_{ik}=-X_{ki}$ can be decomposed as
\be
X_{ik}=X_{\bi\bk}+(X_{\bi\perp}n_k-X_{\bk\perp}n_i)\, .
\ee

The parallel tensors, like $X_{\bi\bk}$, lie in $\S$, and can be further
decomposed into the irreducible parts with respect to the spatial
rotations:
\bsubeq
\be
X_{\bi\bk}=\irr{T}{X}_{\bi\bk}+\irr{A}{X}_{\bi\bk}
          +\frac{1}{2}\eta_{\bi\bk}\irr{S}{X}\, ,
\ee
where
\bea
&&\irr{T}{X}_{\bi\bk}:=X_{(\bi\bk)}
                 -\frac{1}{2}\eta_{\bi\bk}X^\bm{_\bm}\, ,  \nn\\
&&\irr{A}{X}_{\bi\bk}:=X_{[\bi\bk]}\, , \qquad
  \irr{S}{X}:=X^\bm{_\bm}\, .                              \nn
\eea
As a consequence, the product $X^{\bi\bk}Y_{\bi\bk}$ is given by
\be
X^{\bi\bk}Y_{\bi\bk}=\irr{T}{X}^{\bi\bk}\,\irr{T}{Y}_{\bi\bk}
   +\irr{A}{X}^{\bi\bk}\,\irr{A}{Y}_{\bi\bk}
   +\frac{1}{2}\irr{S}{X}\,\irr{S}{Y}\, .
\ee
\esubeq

For a tensor $\Phi_{\bi\bj\bk}=-\Phi_{\bj\bi\bk}$, the pseudoscalar
$(\ve^{\bi\bj\bk}\Phi_{\bi\bj\bk})$ and traceless-symmetric piece
$(\Phi_{\bi(\bj\bk)}-\text{traces})$ identically vanish, so that the only
nontrivial piece is the vector $\irr{V}{\Phi}_\bi:=\Phi_{\bi\bk}{^\bk}$:
\be
\Phi_{\bi\bj\bk}=2\eta_{[\bj\bk}\irr{V}{\Phi}_{\bi]}\, ,\qquad
\Phi^{\bi\bj\bk}Q_{\bi\bj\bk}=2\,\irr{V}{\Phi}^\bi\,\irr{V}{Q}_\bi\, .
\ee

\prg{(2)} These results can be now used to find the Dirac-ADM form of the
Hamiltonian. Starting with the torsion sector, we use the formula
$T=\fT+\cT$ to rewrite $\cL_{T^2}$ in the form
\bea
\cL_{T^2}&=&\frac{1}{4}\cH^{ijk}(\fT)\fT_{ijk}
   +\frac{1}{4}\cH^{ijk}(\cT)\fT_{ijk}
   +\frac{1}{4}\cH^{ijk}(T)\cT_{ijk}                       \nn\\
&=&\cL_T(\fT)+\frac{\hpi^{i\bj}}{J}T_{i\perp\bj}
   -\frac{1}{2}\phi^{i\bj}T_{i\perp\bj}\, ,                \nn
\eea
which yields
\bsubeq
\be
\cH_\perp^T=\frac{1}{2}J\phi^{i\bj}T_{i\perp\bj}
   -J\cL_{T^2}(\fT)-n^i\nab_\a\pi_i{^\a}\,.
\ee
Then, the irreducible decomposition
\be
\phi^{i\bj}T_{i\perp\bj}=\phi^{\perp\bj}\,T_{\perp\perp\bj}
  +\irr{A}{\phi}^{\bi\bj}\,T_{[\bi\perp\bj]}
  +\irr{T}{\phi}^{\bi\bj}\,T_{(\bi\perp\bj)}
  +\frac{1}{2}\irr{S}{\phi}\,T^\bi{}_{\perp\bi}\, ,
\ee
\esubeq
in conjunction with \eq{4.3}, leads to \eq{4.8}.

Similar calculations for the curvature part yield
\bsubeq
\be
\cH^R_\perp=\frac{1}{4}J\Phi^{ij\bk}R_{ij\perp\bk}
               -J\cL_{R^2}(\fR)+aJ\fR\, .
\ee
Then, the irreducible decompositions
\bea
\Phi^{ij\bk}R_{ij\perp\bk}&=&2\Phi^{\perp\bj\bk}R_{\perp\bj\perp\bk}
   +2\irr{V}{\Phi}^\bi R_{\bi\bk\perp}{^\bk} \, ,   \nn\\
2\Phi^{\perp\bj\bk}R_{\perp\bj\perp\bk}&=&
   2\left(\irr{A}{\Phi}^{\perp\bj\bk}\irr{A}{R}_{\bj\perp\bk\perp}
   +\irr{T}{\Phi}^{\perp\bj\bk}\,\irr{T}{R}_{\bj\perp\bk\perp}
   +\frac{1}{2}\irr{S}{\Phi}^\perp R^\bk{}_{\perp\bk\perp}\right),
\eea
\esubeq
combined with \eq{4.5}, lead directly to \eq{4.9}.

\section{Constraint bifurcation in the spin-\mb{0^+} sector}
\setcounter{equation}{0}

In this appendix, we study the phenomenon of constraint bifurcation in the
spin-$0^+$ sector, determined by the critical condition $\Om=0$.

\prg{1.} We start our discussion by writing the field equations for the
spin-$0^+$ sector:
\bea
&&2a_2\eta_{i[j}\nab^k
V_{k]}+2\left(a-\frac{b_6}{6}R\right)G_{ji}
  -\eta_{ij}\left(\frac{a_2}2V^2+\frac{b_6}{12}R^2+2\L_0\right)=0\,,
                                                           \lab{B.1}\\
&&\left(a-\frac{b_6}{6}R\right)T_{ijk}+a_2\eta_{i[j}V_{k]}
  +\frac{b_6}{3}\eta_{i[j}\nab_{k]}R=0\, ,                 \lab{B.2}
\eea
where $V^2=V_kV^k$. The content of these equations can be expressed in
terms of their irreducible components. For the first equation, we find
\bsubeq
\bea
&&-a_2\nabla_{[i}V_{j]}+2\left(a-\frac{b_6}6R\right)R_{[ji]}=0\,,\lab{B.3a}\\
&&-a_2\left(\nab_{(i}V_{j)}-\frac{1}{3}\eta_{ij}\s\right)
  +2\left(a-\frac{b_6}{6}R\right)
           \left(R_{(ji)}-\frac{1}{3}\eta_{ij}R\right)=0\,,\lab{B.3b}\\
&&-2a_2\s+\frac{3}{2}a_2 V^2+aR+\frac{b_6}{12}R^2+6\L_0=0\,,\lab{B.3c}
\eea
\esubeq
where $\s:=\nab_i V^i$. The irreducible components of the second equation
are:
\bsubeq\lab{B.4}
\bea
&&\left(a-\frac{b_6}6R\right)\cA=0\,,                      \lab{B.4a}\\
&&\left(a-\frac{b_6}6R\right)\irr{T}{T}_{ijk}=0\, ,        \lab{B.4b}\\
&&\left(a-\frac{b_6}6R+a_2\right)V_i+\frac{b_6}3\nab_i R=0\,.\lab{B.4c}
\eea
\esubeq

\prg{2.} Now, we focus our attention on the factor $\Om=W-a_2$ in $\D^+$.
Its dynamical evolution is determined by Eq. \eq{B.4c}, which can be
written in the form
\be
-\Om V_k+2\pd_k\Om\approx 0\, ,                            \lab{B.5}
\ee
Note that this equation is an extension of Eq. \eq{5.13} from $\S$ to the
whole spacetime $\cM$.

The spacetime continuum $\cM$ on which 3D PGT lives is a differentiable
manifold with topology $\cM=R\times\S$, where $R$ corresponds to time, and
$\S$ to the spatial section of $\cM$. Let us now assume that: (i) $\Om$
vanishes at some point $x=a$ in $\cM$, (ii) $\Om$ is an infinitely
differentiable function on $\cM$, and (iii) $V_k$ and all its derivatives
are finite at $x=a$. Then, one can notice that \eq{B.5} implies
$\pd_k\Om=0$ at $x=a$. In the next step, we apply the differential
operator $\pd_{k_1}$ to \eq{B.5} and conclude that $\pd_{k_1}\pd_k\Om=0$
at $x=a$. Continuing this procedure, we eventually conclude that for every
$n$, $\pd_{k_n}\cdots\pd_{k_1}\pd_k\Om=0$ at $x=a$. In general, the
behavior of $\Om$ on the whole $\cM$ is not determined by its properties at
a single point. However, if (iv) $\Om$ is an analytic function on $\cM$, its
Taylor expansion around $x=a$ implies that $\Om=0$ on the whole $\cM$.

The result obtained can be formulated in a more useful form: if there is
at least one point in $\cM$ at which $\Om\ne 0$, then $\Om\ne 0$ on the
whole $\cM$. Thus, by choosing the initial data so that $\Om\ne 0$ at
$x^0=0$, it follows that $\Om$ stays nonvanishing for any $x^0>0$. In
other words, for a suitable choice of initial data, the configuration
$\Om=0$ is kind of a barrier that the system cannot cross during its
dynamical evolution. Moreover, since $\Om$ is a continuous function, it
has a definite sign for any $x^0>0$.

\section{The linearized spin-$0^+$ sector}
\setcounter{equation}{0}

In the weak-field approximation, the primary ICs of the spin-$0^+$ sector
take the form
\bea
&&\irr{A}{\tilde\phi}_{\bi\bj}:=
  \frac{\irr{A}{\tilde\hpi}_{\bi\bj}}{\bar J}\approx 0\,,\qquad
\irr{T}{\tilde\phi}_{\bi\bj}:=
  \frac{\irr{T}{\tilde\hpi}_{\bi\bj}}{\bar J}\approx 0\,,  \nn\\
&&\irr{A}{\tilde\Phi}_{\perp\bi\bj}:=
  \frac{{}^A\tilde\hPi_{\perp\bi\bj}}{\bar J}
  -2(a+qb_6)\tilde b_{[\bi\bj]}\approx 0\,,                \nn\\
&&\irr{T}{\tilde\Phi}_{\perp\bi\bj}:=
  \frac{\irr{T}{\hPi}_{\perp\bi\bj}}{\bar J}
  -2(a+qb_6){}^T\tilde b_{\bi\bj}\approx 0\,,              \nn\\
&&\irr{V}{\tilde\Phi}_\bi:=
  \frac{\irr{V}{\tilde\hPi}_{\bi}}{\bar J}
  -2(a+qb_6)\tilde b_{\perp\bi}\approx 0\,,                \lab{C.1}
\eea
and the secondary Hamiltonian constraints are given by
\bsubeq
\bea
&&\tilde\cH_\perp=\bar J\left(2\frac{{}^S\tilde\Pi_\perp}{\bar J}
  -(a+qb_6)(\tR^{\bi\bj}{}_{\bi\bj}-4b^{\bi}{_\bi})\right)
  -\bar n_i\bar\nab_\a\tilde\pi^{i\a}\,,                   \\
&&\tilde\cH_\a=p\ve_{ijk}\bar b^j{_\a}\tilde\hpi_i{^\bk}
  -\bar b^i{_\a}\bar\nab_\b\tilde\pi_i{^\b}
  -2(a+b_6q)\tilde R^0{_\b}\,,                             \\
&&\tilde\cH_{\bi\bj}\approx\frac{{}^A\hpi_{\bi\bj}}{J}
  -(a+qb_6)\tT_{\perp\bi\bj}
  +p\ve_{\perp\bi\bj}\left(\frac{{}^S\tilde\hPi_\perp}{4\bar J}
  +\frac12(a+qb_6)\tb^{\bk}{_\bk}\right)\approx 0\,,       \\
&&\tilde\cH_{\perp\bi}\approx\frac{\tilde\hpi_{\perp\bi}}{\bar J}
  +2(a+qb_6)\tilde T^\bk{}_{\bk\bi}
  +\nab_\bi\left(\frac{{}^S\tilde\hPi_\perp}{2\bar J}
  +(a+qb_6)\tb^\bk{_\bk}\right)\approx 0\, .               \lab{C.2d}
\eea
\esubeq
The consistency condition of $\irr{V}{\tilde\Phi}_\bi$ can be expressed in
the form
\be
\tilde\chi_\bi=\frac{1}{2}\tilde\cH_{\perp\bi}
  -\frac{a+qb_6+a_2}{a_2}\frac{\tilde\hpi_{\perp\bi}}{\bar J}\approx
  \frac{a+qb_6+a_2}{a_2}\frac{\tilde\hpi_{\perp\bi}}{\bar J}\,.\lab{C.3}
\ee
For $a+qb_6\neq 0$ and $a+qb_6+a_2\neq 0$, the type and the number of
constraints remains the same as in the full non-linear theory, and we have
the canonical stability under linearization.

\prg{The case \mb{a+qb_6=0}.} In this case, the analysis depends on the
value of $p$.

(i) for $p\neq 0$, the six secondary constraints
$\cH_M=(\tilde\cH_\perp,\tilde\cH_\a,\tilde\cH_{\bi\bj},\tilde\cH_{\perp\bi})$,
in conjunction with $\irr{V}{\tilde\Phi}_\bi\approx 0$ take, respectively,
the following form:
\bea
\irr{S}{\tilde\hpi}\approx 0\,,\qquad 0\approx 0\,,\qquad
\irr{S}{\tilde\hPi}_\perp\approx 0\,,\qquad \tilde\hpi_{\perp\bi}\approx 0\,.
\eea
Thus, $\tilde\chi_\bi$ and $\tilde\cH_\a$ are identically satisfied, and
there are no SC constraints, $N_2=0$. Hence, the number of FC
constraints is $N_1=6+6-2+8=18$ and consequently, there are no propagating
modes: $N^*=2\times 18-2\times 18-0=0$.

(ii) For $p=0$, the constraints $\cH_M$, in conjunction with
$\irr{V}{\tilde\Phi}_\bi\approx 0$, read:
\bea
&&2{}^S\tilde\Pi_\perp
  -\bar n_i\bar\nab_\a\tilde\pi^{i\a}\approx 0\, ,\qquad
\bar b^i{_\a}\bar\nab_\b\tilde\pi_i{^\b}\approx 0\,,       \nn\\
&&0\approx 0\,,\qquad \frac{\tilde\hpi_{\perp\bi}}{\bar J}
  +\nab_\bi\left(\frac{{}^S\tilde\hPi_\perp}{2\bar J}\right)\approx 0\,.
\eea
Taking into account the form of $\tilde\chi_\bi$, the set $\cH_M$ reduces
to:
$$
\tilde\hpi_{\perp\bi}\approx 0\,,\qquad {}^S\tilde\hpi\approx 0\,,
\qquad {}^S\tilde\hPi_{\perp}\approx 0\,.
$$
Thus, we again have $N_1=18$, $N_2=0$, and $N^*=0$.

\prg{The case \mb{a+qb_6+a_2=0}.} Compared to the generic case, this
condition induces the following change: Eq. \eq{C.3} implies that
$\tilde\chi_\bi$ is identically satisfied, whereas $\irr{V}{\Phi}_\bk$
becomes FC. Thus, $N_1=6+6+2=14$, $N_2=10-4=6$, and consequently, $N^*=2$.

\section{The algebra of ICs in the spin-\mb{0^-} sector}
\setcounter{equation}{0}

The non-trivial PBs between the primary ICs
$Y_M=(\irr{S}{\phi},\irr{T}{\phi},\phi_{\perp,\bk}, \irr{S}{\Phi},
\irr{T}{\Phi})$ in the spin-$0^-$ sector read:
\bea
&&\{\irr{S}{\phi},\irr{S}{\Phi}_\perp\}\approx-\frac{4a}{J}\d\,,\nn\\
&&\{\irr{T}{\phi}_{\bi\bj},\irr{T}{\Phi}_\perp{}^{\bm\bn}\}\approx
  -\frac{1}{J^2}\left[\d_{(\bi}{}^{(\bm}
  \irr{A}{\hPi}_{\perp\bj)}{}^{\bn)}
  -2a\d_{(\bi}{}^\bm\d_{\bj)}{}^\bn\right]\d\, ,           \nn\\
&&\{\phi_{\perp\bi},\phi_{\perp\bj}\}
  \approx \frac{2}{J^2}\irr{A}\hpi_{\bi\bj}\d\, ,          \nn\\
&&\{\phi_{\perp\bi},\irr{S}\Phi_\perp\}
  \approx \frac1{J^2}\irr{V}\hPi_{\bi}\d\, ,               \nn\\
&&\{\phi_{\perp\bi},\irr{T}\Phi_{\perp\bm\bn}\}
  \approx \frac{1}{J^2}\left(\frac{1}{2}\eta_{\bm\bn}\irr{V}\hPi_{\bi}
  -\eta_{\bi(\bm}\irr{V}\hPi_{\bn)}-4aJn_{(m}\eta_{\bi n)}\right)\d\,.\lab{D.1}
\eea
Calculating the determinant of the $8\times 8$ matrix
$\D^-_{MN}=\{Y_M,Y_N\}$,
$$
\D^-=\left|\ba{ccccc}
0&0&-\dis\frac{4a}J&0&0\\
0 & \{\irr{T}{\phi}_{\bi\bj},\irr{T}{\phi}_{\bm\bn}\} & 0 & 0 &
   \{\irr{T}{\phi}_{\bi\bj},\irr{T}{\Phi}_\perp{}^{\bm\bn}\}  \\[3pt]
0 & 0 & \{\phi_{\perp\bi},\phi_{\perp\bj}\} &
   \{\phi_{\perp\bi},\irr{S}{\Phi}_\perp\} &
   \{\phi_{\perp\bi},\irr{T}{\Phi}_\perp{}^{\bm\bn}\}         \\[2pt]
\dis\frac{4a}{J} & 0 & -\{\phi_{\perp\bi},\irr{S}{\Phi}_\perp\} & 0 & 0 \\[2pt]
0 & -\{\irr{T}{\phi}_{\bi\bj},\irr{T}{\Phi}_\perp{}^{\bm\bn}\} &
  -\{\phi_{\perp\bi},\irr{T}{\Phi}_\perp{}^{\bm\bn}\} & 0 & 0
\ea
\right|\nn
$$
one obtains the result displayed in Eq. \eq{6.7}.

\section{On the condition \mb{\irr{A}\hpi_{\bi\bj}=0}}
\setcounter{equation}{0}

In this appendix, we wish to clarify the phenomenon of constraint
bifurcation in the spin-$0^-$ sector, where the field equations take the
form:
\bea
&&2a_3\ve_{ijk}(\nab^k-V^k)\cA
  +\eta_{ij}(a_3\cA^2-b_5 R_{[ij]}R^{[ij]}-2\L_0)          \nn\\
&&\qquad +\frac{8a_3}3\cA\ve_{(imn}t_{j)}{}^{mn}\cA
  +2aG_{ji}+2b_5R_{[in]}G_j{^n}=0\, ,                      \lab{E.1}\\
&&T^i{}_{mn}\left(a\eta_{ik}+b_5R_{[ik]}\right)+b_5\nab_{[m}(R_{n]k}
  -R_{kn]})+2a_3\ve_{kmn}\cA=0\, .                         \lab{E.2}
\eea
Consider now the solutions for which $\irr{A}\hpi_{\bi\bj}=0$, or
equivalently, $\cA=0$. Although the condition $\cA=0$ is imposed from
outside, it is, essentially, an \emph{additional} constraint as compared
to the generic case. Hence, using not only $\cA=0$ but also $\nab^0\cA=0$,
which ensures dynamical consistency of $\cA=0$, the field equations reduce
to
\bsubeq
\bea
&&2aG_{ji}+2b_5R_{[in]}G_j{^n}
  -\eta_{ij}(b_5 R_{[ij]}R^{[ij]}+2\L_0)=0\, ,             \lab{E.3a}\\
&&T^i{}_{mn}\left(a\eta_{ik}+b_5R_{[ik]}\right)
                 +b_5\nab_{[m}(R_{n]k}-R_{kn]})=0\, .      \lab{E.3b}
\eea
\esubeq

In what follows, we focus our attention on the first first equation
\eq{E.3a}. We begin by noting that its components $i,j=\perp\perp$ and
$\bi\perp$, rewritten in the form
\bea
&&-aR^{\bi\bj}{}_{\bi\bj}+2\L_0-\frac{\irr{V}\hPi^{\bi}}JR_{\perp\bi}
  -\frac{1}{4b_5J^2}\left(2\irr{V}\hPi_{\bi}\irr{V}\hPi^{\bi}
  +\irr{A}\hPi_{\perp\bi\bj}\irr{A}\hPi_{\perp}{}^{\bi\bj}\right)=0\,,\\
&&2aR_{\perp\bi}+\frac{\irr{V}\hPi_\bi}JG_{\perp\perp}
  +\frac{\irr{A}\hPi_{\perp\bi\bj}}{J}R_{\perp}{^\bj}=0\, ,
\eea
represent the secondary FC constraints $\cH_\perp$ and
$\cH_\bi:=h_\bi{^\a}\cH_\a$. Next, consider the components $[\bi,\bj]$ and
$[\perp,\bi]$:
\bea
&&\left(2a+b_5R_{\perp\perp}\right)\irr{A}\hPi_{\perp\bi\bj}
   +2b_5\irr{V}\hPi_{[\bi}R_{\perp\bj]}=0\, ,              \lab{E.6}\\
&&2a\irr{V}\hPi_\bi-b_5\left(\irr{V}\hPi_\bi G_{\perp\perp}
  +\irr{A}\hPi_{\perp\bi\bj}R_\perp{^\bj}
  +\irr{V}\hPi_{\bj}G_\bi{^\bj}\right)=0\, .               \lab{E.7}
\eea
Taking into account the relations
\bea
&&aR=\frac{1}{4b_5J^2}\left(2\irr{V}\hPi_{\bi}\irr{V}\hPi^{\bi}
  +\irr{A}\hPi_{\perp\bi\bj}\irr{A}\hPi_{\perp}{}^{\bi\bj}\right)-6\L_0\, ,
                                                           \nn\\
&&G_{\perp\perp}=-\frac 12 R^{\bi\bj}{}_{\bi\bj}\,,        \nn
\eea
it follows that \eq{E.6} is a new constraint. Finally, the $\bi\bj$
components of \eq{E.3a} read:
\bea
\left(2a\eta_{\bi\bk}+\frac{\irr{A}\hPi_{\perp\bi\bk}}J\right)G_\bj{^\bk}
  -\eta_{\bi\bj}\left[\frac{1}{4b_5J^2}\left(2\irr{V}\hPi_{\bm}\irr{V}\hPi^{\bm}
  +\irr{A}\hPi_{\perp\bm\bn}\irr{A}\hPi_{\perp}{}^{\bm\bn}\right)
  +2\L_0\right]=0\, .                                      \nn
\eea
This equation can be solved for $G_\bi{^\bk}$ since
$$
\det\left[2a\eta_{\bi\bk}+\frac{\irr{A}\hPi_{\perp\bi\bk}}J\right]
 =4a^2+\frac{\irr{A}\hPi_{\perp\bm\bn}\irr{A}\hPi_{\perp}{}^{\bm\bn}}{2J^2}>0\,.
$$
Thus, $G_\bi{^\bj}$ can be expressed in terms of the phase-space
variables, and consequently, \eq{E.7} is also a constraint. Hence,
equations \eq{E.6}, \eq{E.7} and $\irr{A}\hpi_{\bi\bj}=0$  describe
\emph{four} additional SC constraints (if any of these were FC, the number
of DoF would be negative), which eliminate the two propagating modes of
the generic case, so that $N^*=0$.

\section{The linearized spin-\mb{0^-} sector}
\setcounter{equation}{0}

In this Appendix, we present the canonical structure of the linearized
spin-$0^-$ sector around the maximally symmetric background. We start by
noting that
$$
\frac{\irr{A}{\bar\hpi}_{\bi\bj}}{J}=-2a_3\ve_{\perp\bi\bj}\bar\cA\, ,
$$
where $\bar\cA=p$. Then, for $p\ne 0$,  Eq. \eq{6.7} implies that the
determinant $\bar\D^-$ is positive definite, so that the canonical
structure remains the same as before linearization, see Section 6.
Moreover, in that case the spin-$0^-$ mode is massless, see \eq{3.8c}.

To see what happens in the complementary case $p=0$ (the spin-$0^-$ mode
is either massive or massless), we start with
\bea
&&\tilde\pi_i{}^\a=2a_3\ve^{0\a\b}\bar b_{i\b}\tilde\cA\, ,\nn\\
&&\tilde\Pi_{ij}{}^\a=-2\ve_{ijk}\left[a\ve^{0\a\b}\tb^k{_\b}
  +b_5\bar b^k{_\r}\ve^{[\a\n\r}
  \left(\tR_\n{}^{0]}-\tR^{0]}{}_\n\right)\right]\, ,      \nn
\eea
and find the following primary ICs:
\bea
&&\irr{S}{\tilde\phi}:=
  \frac{\irr{S}{\tilde\hpi}}{\bar J}\approx 0\, ,\qquad
\irr{T}{\tilde\phi}_{\bi\bj}:=
  \frac{\irr{T}{\tilde\hpi}_{\bi\bj}}{\bar J}\approx 0\, ,\qquad
\tilde\phi_{\perp\bi}:=
  \frac{\tilde\hpi_{\perp\bi}}{\bar J}\approx 0\,,         \nn\\
&&\irr{S}{\tilde\Phi}_\perp:=
  \frac{\irr{S}{\tilde\hPi}_\perp}{\bar J}
  +2a\tb^\bi{_\bi}\approx 0\,,\qquad
\irr{T}{\tilde\Phi}_{\perp\bi\bj}:=
  \frac{\irr{T}{\tilde\hPi}_{\perp\bi\bj}}{\bar J}
  -2a{}^T\tb_{\bi\bj}\approx 0\, .                         \lab{F.1}
\eea
The only nontrivial  PBs between the primary ICs are:
\bea
&&\{\irr{S}{\tilde\Phi}_{\perp},\irr{S}{\tilde\phi}\}
  \approx\frac{4a}{\bar J}\d\, ,                           \nn\\
&&\{\irr{T}{\tilde\Phi}_{\perp\bi\bj},\irr{T}{\tilde\phi}^{\bm\bn}\}
  \approx-\frac{2a}{\bar J}\d_{(\bi}{}^{\bm}\d_{\bj)}{}^\bn\d\,.\lab{F.2}
\eea
The secondary constraints $\tilde\cH_i$ and $\tilde\cH_{ij}$ read:
\bea
&&\tilde\cH_\perp=\bar J\left[2\frac{\irr{S}\tilde\Pi_\perp}{\bar J}
  -a(\tR^{\bi\bj}{}_{\bi\bj}-4b^{\bi}{_\bi})\right]\, ,    \nn\\
&&\tilde\cH_\a=\bar b^\bi{_\a}\left[-q\left(\irr{V}\tilde\hPi_{\bi}
  -2aJ\tb_{\perp\bi}\right)
 -\bar\nab_\b\tilde\pi_i{^\b}-2a\bar J\tR_{\perp\bi}\right]\,,\nn\\
&&\tilde\cH_{\bi\bj}\approx
   a\tT_{\perp\bi\bj}+\frac{\irr{A}\tilde\hpi_{\bi\bj}}{2\bar J}
  +\nab_{[\bi}\left(\frac{\irr{V}\tilde\hPi_{\bj]}}{\bJ}
  -2a\tb_{\perp\bj]}\right)\, ,                            \nn\\
&&\tilde\cH_{\perp\bi}\approx
  a\tilde\fV_\bi+\frac{1}{2}\bar\nab^\bj\left(
  \frac{\irr{A}\tilde\hPi_{\perp\bi\bj}}{\bar J}-2a\tb_{[\bi\bj]}\right)\,,
\eea
Moreover, $\tilde\cH_\a$ can be used to find
$\tilde\cH_\bi={\bar h_\bi{^\a}\tilde\cH_\a}/{\bar J}$:
\be
\tilde\cH_\bi\approx
  -q\left(\frac{\irr{V}\tilde\hPi_{\bi}}{\bar J}
                  -2a\tb_{\perp\bi}\right)
  +\bar\nab^\bj\frac{\irr{A}\tilde\hpi_{\bj\bi}}{\bar J}
  -2a\tR_{\perp\bi}\, .                                    \nn
\ee

According to \eq{F.2}, the consistency of the primary ICs
$(\irr{S}{\phi},\irr{T}{\phi};\irr{S}{\Phi},\irr{T}{\Phi})$ results in the
determination of the multipliers
$(\irr{S}{u},\irr{T}{u};\irr{S}{v},\irr{T}{v})$, whereas the consistency
of $\tilde\phi_{\perp\bi}$ yields a new, secondary IC:
\be
\tilde\chi_\bi=\bar\nab^\bj\frac{\irr{A}\tilde\hpi_{\bj\bi}}{\bar J}+
\frac{1}{b_5}(a-qb_5)\left(\frac{\irr{V}\tilde\hPi_\bi}{\bar J}
  -2a\tilde b_{\perp\bi}\right)\approx 0\, .
\ee
The PB of $\tilde\phi_{\perp\bi}$ with its own (modified) secondary pair
$\tilde\chi'_\bj:=\tilde\chi_\bj-\tilde\cH_\bj$ reads:
\be
\{\tilde\phi_{\perp\bi},\tilde\chi'_{\bj}\}=
       \frac{2a^2}{b_5\bar J}\eta_{\bi\bj}\d\, .           \lab{F.5}
\ee
Thus, the consistency condition of $\tilde\chi'_\bj$ leads to the
determination of the multiplier $u_{\perp\bj}\,$.

According to Eqs. \eq{F.2} and \eq{F.5}, the ten ICs $\tilde X_A=
\{\irr{S}{\tilde\phi},\irr{S}{\tilde\Phi}_{\perp},\irr{T}{\tilde\phi}_{\bi\bj},
\irr{T}{\tilde\Phi}_{\perp\bi\bj},\tilde\phi_{\perp\bi},\tilde\chi'_\bi\}$
are SC. Hence, $N=18$, $N_1=12$, $N_2=10$, so that $N^*=2$ (one Lagrangian
DoF).


\end{document}